\documentclass[a4paper,11pt]{article}
\pdfoutput=1 
\usepackage{savesym}
\usepackage[makeroom]{cancel}
\usepackage{slashed, tikz}
\usepackage{graphicx,array}
\usepackage{booktabs}
\usepackage{dsfont}
\usepackage{stackrel}
\usepackage{setspace}
\usepackage[normalem]{ulem}
\usepackage[T1]{fontenc} 
\usepackage{xcolor}
\usepackage{changepage}
\usepackage[most]{tcolorbox}
\usepackage{float}
\usepackage{caption}
\usepackage{subfigure}
\usepackage{jheppub} 
\usepackage{stackrel}

\makeatletter\g@addto@macro\bfseries{\boldmath}\makeatother
\DeclareMathSymbol{\shortminus}{\mathbin}{AMSa}{"39}
\def\figureautorefname~#1\null{fig.\,#1\null}
\def\equationautorefname~#1\null{eq.\,(#1)\null}

\makeatletter
\DeclareFontFamily{OMX}{MnSymbolE}{}
\DeclareSymbolFont{MnLargeSymbols}{OMX}{MnSymbolE}{m}{n}
\SetSymbolFont{MnLargeSymbols}{bold}{OMX}{MnSymbolE}{b}{n}
\DeclareFontShape{OMX}{MnSymbolE}{m}{n}{
    <-6>  MnSymbolE5
   <6-7>  MnSymbolE6
   <7-8>  MnSymbolE7
   <8-9>  MnSymbolE8
   <9-10> MnSymbolE9
  <10-12> MnSymbolE10
  <12->   MnSymbolE12
}{}
\DeclareFontShape{OMX}{MnSymbolE}{b}{n}{
    <-6>  MnSymbolE-Bold5
   <6-7>  MnSymbolE-Bold6
   <7-8>  MnSymbolE-Bold7
   <8-9>  MnSymbolE-Bold8
   <9-10> MnSymbolE-Bold9
  <10-12> MnSymbolE-Bold10
  <12->   MnSymbolE-Bold12
}{}

\let\llangle\@undefined
\let\rrangle\@undefined
\DeclareMathDelimiter{\llangle}{\mathopen}%
                     {MnLargeSymbols}{'164}{MnLargeSymbols}{'164}
\DeclareMathDelimiter{\rrangle}{\mathclose}%
                     {MnLargeSymbols}{'171}{MnLargeSymbols}{'171}
\makeatother

\definecolor{orange}{rgb}{1,0.5,0}
\definecolor{brown}{rgb}{0.59, 0.29, 0.0}

\definecolor{note_fontcolor}{rgb}{0.80078125, 0.80078125, 0.80078125}

\newtcbox{\mymath}[1][]{%
    nobeforeafter, math upper, tcbox raise base,
    enhanced, colframe=blue!30!black,
    colback=white, boxrule=1pt,
    #1}

\def\beq{\begin{equation}}
\def\eeq{\end{equation}}
\def\bea{\begin{eqnarray}}
\def\eea{\end{eqnarray}}

\makeatletter
\newcommand*{\Relbarfill@}{\arrowfill@\Relbar\Relbar\Relbar}
\newcommand*{\xeq}[2][]{\ext@arrow 0055\Relbarfill@{#1}{#2}}
\makeatother

\newcommand{\colvec}[1]{\left(\begin{array}{c}#1\end{array}\right)}

\newcommand{\colmatth}[1]{\left(\begin{array}{ccc}#1\end{array}\right)}

\definecolor{MHBlue}{RGB}{156, 156, 252}

\def\lpar#1#2#3#4{\rlap{\raise#3\hbox{$\hskip#4#1\left\{\mbox{\phantom{\rule[0mm]{0mm}{#2}}}\right.$}}}
\def\rpar#1#2#3#4{\rlap{\raise#3\hbox{$\hskip#4\left\}#1\mbox{\phantom{\rule[0mm]{0mm}{#2}}}\right.$}}}
\renewcommand{\subsubsection}[1]{\addtocounter{subsubsection}{1}
\par\nobreak
\medskip
\nobreak
\noindent{\it \thesubsubsection.  #1 }
\par\nobreak\medskip\nobreak}

\definecolor{MHPink}{RGB}{255, 227, 253}
\definecolor{MHOrange}{RGB}{255, 232, 194}
\definecolor{MHCyan}{RGB}{209, 241, 255}
\definecolor{MHGreen}{RGB}{222, 255, 231}

\title{\boldmath Dressed vs. Pairwise States, and  
the Geometric Phase of Monopoles and Charges}

\author[a]{Csaba Cs\'aki,}
\affiliation[a]{Department of Physics, LEPP, Cornell University, Ithaca, NY 14853, USA}
\emailAdd{csaki@cornell.edu}
\author[a,b,c]{Zi-Yu Dong,}
\affiliation[b]{CAS Key Laboratory of Theoretical Physics, Institute of Theoretical Physics, Chinese Academy of Sciences, Beijing 100190, China.}
\affiliation[c]{School of Physical Sciences, University of Chinese Academy of Sciences, Beijing 100190, P. R. China.}
\emailAdd{zd79@cornell.edu}
\author[d]{Ofri Telem,}
\affiliation[d]{Racah Institute of Physics, Hebrew University of Jerusalem, Jerusalem 91904, Israel}
\emailAdd{t10ofrit@gmail.com}
\author[e]{John Terning}
\emailAdd{jterning@gmail.com}
\affiliation[e]{QMAP, Department of Physics, University of California, Davis, CA 95616, USA}
\author[f]{and Shimon Yankielowicz}
\affiliation[f]{The Raymond and Beverly Sackler School of Physics and Astronomy,
Tel Aviv University, Ramat Aviv 69978, Israel}
\emailAdd{shimonya@tauex.tau.ac.il}

\abstract{We construct the Faddeev-Kulish dressed multiparticle states of electrically and magnetically charged particles, incorporating the effects of real and virtual soft photons. We calculate the  properties of such dressed states under Lorentz transformations, and find that they can be identified with the pairwise multi-particle states that transform under the pairwise little group. The shifts in the dressing factors under Lorentz transformations are finite and have a simple geometric interpretation. Using the transformation properties of the dressed states we also present a novel, fully quantum field theoretic derivation of the geometric (Berry) phase obtained by an adiabatic rotation of the Dirac string, and also of the Dirac quantization condition. For half integer pairwise helcity, we show that these multiparticle states have flipped spin-statistics, reproducing the surprising fact that fermions can be made out of bosons.}

\begin{document}
\maketitle
  
\section{Introduction}

In his seminal paper \cite{Dirac:1931kp} from 1931, Dirac initiated the study of quantum electric charges in the presence of magnetic monopoles. In modern language, Dirac's main argument is that one should not be able to detect the magnetic flux carried by the Dirac string by performing an Aharonov-Bohm \cite{Aharonov1959} experiment around it, since the string is purely a gauge artifact. 

Dirac's quantization argument can be restated in terms of the geometric phase of the charge-monopole system. The concept of a geometric phase in quantum physics was first explored by Pancharatnam \cite{Pancharatnam1956} in the context of optics, and by Berry \cite{Berry1984} for general quantum systems. A geometric phase occurs when the Hamiltonian $H(\alpha_i)$ for a quantum system depends on a set of external parameters, $\alpha_i$. If the parameters $\alpha_i$ are varied adiabatically over a closed loop in the parameter space, the eigenstates $\left|n;\alpha_i\right\rangle$ return transformed by an overall phase,
\begin{eqnarray}\label{eq:qsberry}
\left|n;\alpha\right\rangle \rightarrow  e^{i\gamma_{B}}\left|n;\alpha\right\rangle\,,
\end{eqnarray}
with the geometric phase (the Berry phase) $\gamma_{B}$ given by
\begin{eqnarray}\label{eq:gammaberry}
\gamma_{B}=i\int_{C}\,d\alpha_i\,\left\langle n;\alpha\right|\frac{d}{d\alpha_i}\left|n;\alpha\right\rangle\,,
\end{eqnarray}
and $C$ is a closed curve in the parameter space. 

In his original paper \cite{Berry1984}, Berry described the Aharonov-Bohm phase as a special case of a geometric phase: we can take a quantum particle of charge $e$ in a box and consider the Hamiltonian $H(\vec{X})$ of the box as a function of its position $\vec{X}$, were $\vec{A}(\vec{X})$ is a background electromagnetic field (in this case generated by a monopole). Now consider moving the box in space along an infinitesimal closed trajectory, such that the system picks up a geometric phase $\gamma_{B}=e\,\int_C\,\vec{A}\cdot d\vec{X}=e\Phi_m$, proportional to the overall magnetic flux captured by a Stokes surface linked to the trajectory $C$ of the box. In this context, the geometric phase is the same as the Aharonov-Bohm phase \cite{Aharonov1959}  $\gamma_{AB}$ generated by the flux $\Phi_m$. But for an infinitesimal trajectory, the phase is either $\gamma_{AB}=0$ or $\gamma_{AB}=eg$, depending whether or not the Stokes surface intersects the Dirac string. Since the Dirac string is an unobservable gauge artifact, we arrive at the Dirac quantization condition: 
\begin{eqnarray}\label{eq:DirQuant}
q\equiv\frac{eg}{4\pi}=\frac{n}{2}\,,
\end{eqnarray}
where $n$ is an integer. 

Though a ground-breaking achievement at the time, there is something unsatisfying about the above argument for the geometric phase of a charge-monopole system and the corresponding Dirac quantization. First, it treats the monopole and the charge on unequal footing---the monopole is taken to be a static, classical object, while the charge is fully quantum mechanical. Such an uneven treatment lacks elegance, especially due to the inherent electric-magnetic duality of Maxwell's equations with magnetic charges. Secondly, the geometric phase associated with the Dirac string does not rely on any non-relativistic expansion, hence it should be possible to cast it in the language of Quantum Field Theory (QFT). Despite progress in relating the Aharonov-Bohm phase to the 4D linking number generated by soft-photon resummation \cite{Terning:2018udc},  a complete QFT description of the geometric phase is missing from the literature, even 90 years after Dirac's original paper. Lastly, the geometric phase associated with the Dirac string is intimately related to the extra angular momentum stored in the EM field sourced by the charge and the monopole. 
It is then natural to ask what form this relation takes in a fully quantum-field-theoretic treatment of charge-monopole multiparticle states.

The current work explores the geometric phase associated with  general multiparticle states of relativistic charges and monopoles. As such, it is the first 4D QFT realization of a geometric phase. A fundamental element in our construction is the use of multiparticle states that 
cannot be spanned by tensor products of a finite number of single particle states. Indeed, 
tensor product states
cannot have a geometric phase, since there is no such phase associated with single-particle states.  In order for a geometric phase to appear we need a coherent state with an indefinite number of particles\footnote{Technically coherent states live in a von Neumann space rather than a simple Fock space \cite{Gomez:2018war}.}. 

The multiparticle states associated with dyons, monopoles and charges 
are known to not be tensor product states,
as was shown in detail by some of the present authors 
\cite{Csaki2021,Csaki:2020uun,Csaki2021a}, following the work of Zwanziger \cite{Zwanziger:1972sx}. Under Lorentz transformations, these charge-monopole, or ``pairwise" states transform with an extra ``pairwise helicity'' little group phase
which exactly mirrors the electromagnetic contribution to the angular momentum from the coherent state of soft photons.

The original mathematical definition of these states in \cite{Csaki2021} was purely group theoretical, and did not clarify the dynamics underlying these states. In this paper we present a simple definition of these states as the soft-photon dressed states (also known as Faddeev-Kulish dressed states) of QED with monopoles, aka ``Quantum Electro-Magneto Dynamics" (QEMD). These dressed
states were previously shown to solve the IR problem for QEMD \cite{Blagojevic1981,Blagojevic1982,Choi2020}. Here we go one step further and show that these dressed states are identical to the pairwise states from \cite{Zwanziger:1972sx,Csaki2021,Csaki:2020uun,Csaki2021a}. Moreover, they transform with a geometric phase which leads to half-integer Dirac quantization and, for half integer pairwise helicity, to an inversion of their spin statistics (e.g. ``making fermions out of bosons'').

This paper is organized as follows; In Section~\ref{sec:summary} we present a concise summary of the results derived in this paper. In Section~\ref{sec:pairwisereview} we briefly review the construction of pairwise multiparticle states, with a special emphasis on the full functional form of the pairwise little group phase $\varphi_{LG}$ and its relation to the Dirac string direction $n^\mu$. 
In Section~\ref{sec:dressQEMD} we briefly review the Faddeev-Kulish \cite{Kulish1970} dressing of asymptotic states in QED, and generalize it to the case of QEMD \cite{Zwanziger:1970hk}. Our first main result is derived in detail in Section~\ref{sec:dressedaspairwise}, namely that the dressed states of QEMD transform under Lorentz transformation exactly the same way as pairwise states. To do this, we act with the Noether charge $M^{\mu\nu}$ for Lorentz transformations in QEMD on the Faddeev-Kulish dressing via commutation. As a byproduct of our work, we evaluate the shift in the Faddeev-Kulish phase $\Phi_{FK}$ under Lorentz transformations, and show that it is finite (compared to log divergent in pure QED). Furthermore it has a geometric meaning as a dihedral angle between two 3-hyperplanes in 4D. Our second main result of the paper is given in Section 6, where we compute the geometric phase associated with a 2$\pi$ rotations of pairwise/dressed QEMD states. This geometric phase turns out to be $\pm 2\pi \sum q_{ij}$, exactly \textit{half} of the Aharonov-Bohm phase associated with the Dirac string. However the geometric phase we calculate is independent of the direction of the string, and so it is allowed to take the values $0$, $\pm \pi$, $\ldots$ leading to half-integer Dirac quantization. In particular, for half-integer $\sum q_{ij}$, the system flips its spin-statistics, as was shown long ago in the static monopole limit \cite{Schwinger1975,Jackiw:1976xx,Hasenfratz1976,Goldhaber1976,Brandt1978,Wilczek1982,Wilczek1982a}. Finally, in Section~\ref{sec:conc}
we summarize our observations and point out some interesting future directions in other dimensions.

\section{Summary of Results}\label{sec:summary}

Before delving into our detailed derivations, we first present a concise summary of our results. Our main result is that the pairwise states---those multiparticle states that transform under the Lorentz group with an extra pairwise phase, can be identified with the Faddeev-Kulish dressed multiparticle states of QEMD.  The latter states incorporate the effects of long distance interactions due to the exchange and radiation of soft photons. The pairwise states $|p_1,p_2,q_{12}\rangle$, with particle momenta $p_i$ and pairwise helicity given by
\beq
 q_{12}=\frac{e_1g_2-e_2g_1}{4 \pi}~,
 \eeq
transform under Lorentz transformations with a pairwise little group phase 
 \begin{eqnarray}\label{eq:2particlenew}
U\left(\Lambda\right)\left|p_1,p_2;q_{12}\right\rangle = e^{-iq_{12}\varphi_{LG}\left(p_1,p_2,\Lambda\right)}\left|\Lambda p_1,\Lambda p_2;q_{12}\right\rangle\,.
\end{eqnarray}
where the explicit expression for the pairwise phase is 
\begin{eqnarray}~\label{eq:pLGphasenew}
\cos\left[\varphi_{LG}(p_1,p_2,\Lambda)\right]&=&\hat{\epsilon}(p_1,p_2,\Lambda^{-1}n)\cdot\hat{\epsilon}(p_1,p_2,n)\,.
\end{eqnarray}
Here $\epsilon_\mu(a,b,c)\equiv\epsilon_{\mu\nu\rho\sigma}a^\nu b^\rho c^\sigma$ and $\hat{\epsilon}_\mu=\epsilon_\mu/\sqrt{\epsilon\cdot\epsilon}$, while $n$ is an arbitrary 4-vector, identified with the direction of the Dirac string. 

The essence of the Faddeev-Kulish dressing procedure is to modify the interaction picture perturbation theory. The usual assumption that interactions fall off asymptotically is not satisfied in QED or QEMD due to the presence of soft photons mediating a long-range force. Hence the Faddeev-Kulish dressing separates out the asymptotic part (corresponding to the soft photons) of the interactions, removes it from the interaction Hamiltonian and uses it to create the Faddeev-Kulish dressed states.\footnote{For a nice review see \cite{Gaharia:2019xlh}.} 
The resulting Faddeev-Kulish dressed states $|p_1, p_2  \rrangle$ of QEMD are constructed from the traditional Fock-space states as 
\begin{eqnarray}\label{eq:dressedEMnew}
\left|p_1,\ldots,p_f\right\rrangle&=&\mathcal{U}_{QEMD}\left|p_1,\ldots,p_f\right\rangle\,,
\end{eqnarray}
where $\mathcal{U}_{QEMD}$ is the dressing factor built out of the asymptotic interaction potential $V^I_{as} (t) = \lim_{|t|\to \infty} V^I(t)$, which as explained captures the effect that the interactions do not asymptotically vanish due to the presence of soft photons. This dressing can be written as a real dressing factor $R_{FK}$ associated with the generation of a real photon cloud, and a phase $\Phi_{FK}$ associated with virtual photon exchange between asymptotic particles. It has been shown that the real dressing factor is equivalent to the usual Wilson line treatment \cite{Jakob:1990zi}. Together these are defined as 
\begin{eqnarray}\label{eq:UeqEMnew}
\mathcal{U}_{QEMD}&\equiv&\mathcal{T}\,\exp{\left[-i\int_{-\infty}^\infty\,dt\,V^I_{as\,;\,QEMD}(t)\right]}=e^{R_{FK}}\,e^{i\Phi_{FK}}\nonumber\\[5pt]
R_{FK}&=&-i\int_{-\infty}^\infty dt ~V^I_{as\,;\,QEMD}(t)\nonumber\\[5pt]  
\Phi_{FK}&=&\frac{i}{2} \int_{-\infty}^\infty dt_1\int_{-\infty}^{t_1}dt_2~[V^I_{as\,;\,QEMD}(t_1),V^I_{as\,;\,QEMD}(t_2)]\,.
\end{eqnarray}
In the bulk of the paper we explicitly show that the dressed states constructed in this way transform exactly as pairwise states should under a Lorentz transformation: 
 \begin{eqnarray}\label{eq:toshow0new}
U\left[\Lambda\right]\,\left|p_1,\ldots,p_f\right\rrangle=e^{i\Phi_{LG}}\left|\Lambda p_1,\ldots,\Lambda p_f\right\rrangle\,,
\end{eqnarray}
where $\Phi_{LG}\equiv\sum_{l<m}q_{lm}\varphi_{LG}(p_l,p_m,\Lambda)$ and $\varphi_{LG}$. Hence the Faddeev-Kulish dressed states can be identified with pairwise states with definite pairwise helicities. The proof will involve the explicit evaluation of both $R_{FK}$ and $\Phi_{FK}$, as well as their transformations under the Lorentz generators $M^{\mu\nu}$. We find that the $\mathcal{O}(eg)$ shift of $\Phi_{FK}$ under Lorentz transformations is finite and given by 
\begin{eqnarray}\label{eq:QM25new}
\Delta\varphi_{FK}(p_1,p_2,n)&=&2\arccos\left[\hat{\epsilon}(p_1,p_2,\Lambda^{-1}n)\cdot\hat{\epsilon}(p_1,p_2, n)\right]\,.
\end{eqnarray}
which is exactly twice the pairwise little group phase. We will also show that the real dressing factor $R_{FK}$ also contributes to the overall Lorentz rotation of the dressed states, and its effect (expressed via single and double commutators with the Lorentz generator) cancels half of $\Delta \varphi_{FK}$, leading to an overall little group phase for the dressed states which is exactly equal to that of the pairwise states. 

Using the transformation properties of our dressed/pairwise states we can find a novel, fully quantum field theoretic derivation of the geometric (``Berry") phase and Dirac quantization associated with any multiparticle states of dyons, monopoles, and/or charges. To obtain a geometric phase, we consider an adiabatic rotation of the Dirac string $n^\mu$ (which is treated as an unobservable parameter of the Lagrangian). Based on our explicit construction we show that rotating the Dirac string is equivalent to an inverse rotation of the entire state
 \begin{eqnarray}\label{eq:Berry2new}
\left|p_1,\ldots,p_f\right\rrangle_{n(\tau+\delta \tau)}=e^{-\frac{i\delta\tau}{2}\omega_{\mu\nu} \Phi^{\mu\nu}_{LG}}\left|p_1,\ldots,p_f\right\rrangle_{n(\tau)}\, 
\end{eqnarray}   
where $\omega_{\mu\nu}$ are the Lorentz transformation parameters. A $2\pi$ rotation will result in a geometric phase of 
\begin{eqnarray}\label{eq:Berry5new}
\gamma_{B}&=&\pm 2\pi \sum_{l<m}\,q_{lm}\,.
\end{eqnarray}
Requiring that this phase be a multiple of $\pi$ gives the usual Dirac quantization condition. Furthermore, multiparticle states with half integer $\sum_{l<m}\,q_{lm}$ will incur a geometric phase of $\pi$, which flips their statistics.

\section{A Group Theoretical Derivation of the Pairwise Little Group Phase}\label{sec:pairwisereview}

In this section we briefly review the construction of the extra phase appearing in multiparticle states carrying both electric and magnetic charges first obtained by Zwanziger in~\cite{Zwanziger:1972sx} and identified as the pairwise helicity of the pairwise little group in~\cite{Csaki2021,Csaki:2020uun,Csaki2021a}. We will be following Zwanziger's original argument, except for using the center of momentum (COM) frame to define the reference momenta, which makes the concept of pairwise little group much easier to understand. 
 
Consider a multiparticle state consisting of a scalar charge and a scalar monopole. This multiparticle state is labeled by the momenta $p_1,p_2$ of the two particles, as well as their \textit{pairwise helicity} $q_{12}\equiv e_1g_2/4\pi$ \cite{Zwanziger:1972sx,Csaki2021,Csaki:2020uun,Csaki2021a}. Under a Lorentz transformation $\Lambda$, this state transforms in a unitary representation of the Lorentz group:
\begin{eqnarray}\label{eq:2particle}
U\left(\Lambda\right)\left|p_1,p_2;q_{12}\right\rangle = e^{-iq_{12}\varphi_{LG}\left(p_1,p_2,\Lambda\right)}\left|\Lambda p_1,\Lambda p_2;q_{12}\right\rangle\,.
\end{eqnarray}
Note the choice of negative sign in the exponential on the RH side---in the language of \cite{Csaki2021,Csaki:2020uun,Csaki2021a}, this means that out states of the S-matrix approach the state \eqref{eq:2particle} at $t\rightarrow\infty$, while in-states transform with the opposite phase. The phase $\varphi_{LG}\left(p_1,p_2,\Lambda\right)$ is called the \textit{pairwise phase} associated with the momenta $p_1,p_2$ and the Lorentz transformation $\Lambda$. It is given by
 \begin{eqnarray}\label{eq:pairph}
R_z\left[\varphi_{LG}(p_1,p_2,\Lambda)\right]^\mu_{~\nu}\equiv \left[L^{-1}_{\Lambda p}\right]^\mu_{~\rho}\Lambda^\rho_{~\sigma}\left[L_{p}\right]^\sigma_{~\nu}\,.
\end{eqnarray}
The formal definition of electric-magnetic multiparticle states was given in \cite{Csaki:2020uun}, based on the ideas of \cite{Zwanziger:1972sx}. Central to the definition of electric-magnetic multiparticle states is the notion of the \textit{pairwise little group} (pairwise LG). This is the subgroup of Lorentz transformations which keeps two reference momenta fixed. In \cite{Csaki2021}, we defined the reference momenta $k^\mu_1,\,k^\mu_2$ as the center f mass (COM) values of $p^\mu_1,\,p^\mu_2$,
\begin{eqnarray}\label{eq:refmom}
k^1_\mu=\left(E^c_1,0,0,p_c\right)~~~,~~~k^2_\mu=\left(E^c_2,0,0,-p_c\right)\,.
\end{eqnarray}
Note that the COM spatial momentum $p_c$ and the COM energies $E^c_{1,2}$ are all Lorentz invariant, and given by
\begin{eqnarray}
E^c_i=\sqrt{m^2_i+p^2_c}~~~,~~~p_c=\sqrt{\frac{(p_1\cdot p_2)^2-m^2_1m^2_2}{s}}\,.
\end{eqnarray}
where $s=(p_1+p_2)^2$. As a first step to defining the pairwise little group (LG) transformation corresponding to a Lorentz transformation $\Lambda$ and the momenta $p_1,\,p_2$, we define a canonical Lorentz transformation $L_p$ so that
\begin{eqnarray}
p^1_\mu=\left[L_p\right]_\mu^{~\nu}k^1_\nu~~~~,~~~~p^2_\mu=\left[L_p\right]^{~\nu}_{\mu}k^2_\nu\,.
\end{eqnarray}
From \eqref{eq:refmom}, we can infer that
\begin{eqnarray}
\left[L_p\right]^{~0}_{\mu}=\frac{p_\mu^1+p_\mu^2}{\sqrt{s}}~~~~,~~~~\left[L_p\right]^{~3}_{\mu}=\frac{E^c_2\,p_\mu^1-E^c_1\,p_\mu^2}{\sqrt{s}p_c}\,.
\end{eqnarray}
Since the columns of $\left[L_p\right]^{~\nu}_{\mu}$ are orthonormal as 4 vectors, we know that $\left[L_p\right]^{~1,2}_{\mu}$ can be any two orthonormal vectors in the plane perpendicular to $p_1$ and $p_2$. The freedom to choose them is exactly the freedom to multiply $L_p$ on the right by a $U(1)$ pairwise LG rotation which keeps $k_{1,2}$ fixed. We fix this freedom by introducing an arbitrary vector $n^\mu$ (analogous to the Dirac string direction in the field theory language), and defining:
\begin{eqnarray}\label{eq:canonch}
\left[L_p\right]^{~1}_\mu=-i\hat{\epsilon}_\mu(p_1,p_2,\hat{\epsilon}(p_1,p_2,n)),~~~~\left[L_p\right]^{~2}_\mu=i\hat{\epsilon}_\mu(p_1,p_2,n)\,.
\end{eqnarray}
Here $\epsilon_\mu(a,b,c)\equiv\epsilon_{\mu\nu\rho\sigma}a^\nu b^\rho c^\sigma$ and $\hat{\epsilon}_\mu=\epsilon_\mu/\sqrt{\epsilon\cdot\epsilon}$.
Substituting this in the definition of the pairwise phase \eqref{eq:pairph}, we get
 \begin{eqnarray}
\cos\left[\varphi_{LG}(p_1,p_2,\Lambda)\right]&=&\left[L^{-1}_{\Lambda p}\right]^{~\rho}_2\Lambda_\rho^{~\sigma}\left[L_{p}\right]_\sigma^{~2}\nonumber\\[5pt]
\sin\left[\varphi_{LG}(p_1,p_2,\Lambda)\right]&=&\left[L^{-1}_{\Lambda p}\right]_2^{~\rho}\Lambda_\rho^{~\sigma}\left[L_{p}\right]_\sigma^{~1}\,.
\end{eqnarray}
By explicit matrix multiplication, we arrive at \cite{Zwanziger:1972sx}
\begin{eqnarray}~\label{eq:pLGphase}
\cos\left[\varphi_{LG}(p_1,p_2,\Lambda)\right]&=&\hat{\epsilon}(p_1,p_2,\Lambda^{-1}n)\cdot\hat{\epsilon}(p_1,p_2,n)\nonumber\\[5pt]
\sin\left[\varphi_{LG}(p_1,p_2,\Lambda)\right]&=&\hat{\epsilon}(p_1,p_2,\Lambda^{-1}n)\cdot\hat{\epsilon}(p_1,p_2,\hat{\epsilon}(p_1,p_2,n))\,,
\end{eqnarray}
this gives the pairwise LG phase as a function of $p_1,\,p_2,\,\Lambda$, and the arbitrary 4-vector $n$. We note that our choice of canonical Lorentz transformation \eqref{eq:canonch} is only unique up to a $U(1)$ pairwise LG z-rotation, which by definition leaves the reference momenta \eqref{eq:refmom} invariant. Defining a more general rotation
\begin{eqnarray}
L_p\rightarrow L_p R_z\left[\chi(p_1,p_2)\right]\,,
\end{eqnarray}
we have 
\begin{eqnarray}
\varphi_{LG}(p_1,p_2,\Lambda)\rightarrow \varphi_{LG}(p_1,p_2,\Lambda) +\chi(p_1,p_2)-\chi(\Lambda p_1,\Lambda p_2)\,.
\end{eqnarray}
For a multiparticle pairwise state\footnote{Here we consider scalars for simplicity.}, the transformations multiply, and so we have
\begin{eqnarray}\label{eq:QM27}
U(\Lambda)\left|p_1,\ldots,p_f;q_{12},q_{13},\ldots,q_{n-1,n}\right\rangle=e^{i\Phi_{LG}}\left|\Lambda p_1,\ldots,\Lambda p_f;q_{12},q_{13},\ldots,q_{n-1,n}\right\rangle\,.\nonumber\\
\end{eqnarray}
where 
\begin{eqnarray}\label{eq:PhiLG}
\Phi_{LG}\equiv-\sum_{l<m}\,q_{lm}\,\varphi_{LG}(p_l,p_m,\Lambda)\,.
\end{eqnarray}
Note the minus sign here, which comes from\eqref{eq:2particle}. Finally, we can consider the pairwise little group phase associated with an infinitesimal Lorentz transformation. Defining $\Lambda^\mu_\nu=\exp(\delta\tau\omega^\mu_\nu)$ for $\delta\tau$ infinitesimal and $\omega^{\mu\nu}$ an antisymmetric matrix, we can expand to first order in $\delta\tau$ and get
\begin{eqnarray}\label{eq:PhimunuLG}
\Phi_{LG}&\equiv&\frac{\delta\tau}{2}\,\omega_{\mu\nu}\,\varphi^{\mu\nu}_{LG}+\mathcal{O}(\delta\tau^2)\nonumber\\[5pt]
\Phi^{\mu\nu}_{LG}&=&-\sum_{l<m}\,q_{lm}\,\varphi^{\mu\nu}_{LG;lm}\nonumber\\[5pt]
\varphi^{\mu\nu}_{LG;lm}&=&\frac{\tau_{lm}}{\epsilon^2(p_l,p_m,n)}\,n^{[\mu}\epsilon^{\nu]}(p_l,p_m,n)\,.
\end{eqnarray}
Here $\tau_{lm}\equiv\sqrt{(p_l\cdot p_m)^2-m^2_lm^2_m}$.

\section{The Soft Photon Dressed States of Monopole QED}\label{sec:dressQEMD}
\subsection{Dressed States in QED}
The IR problem of QED and its solution via dressed states is a deep and rich topic in QFT, and we will not attempt to review it in detail. For historical background and some current developments, see \cite{Morchio2016,Hirai2019,Frye2019,Hannesdottir:2019opa,Hirai2021} and references within. For the purposes of this paper, we will focus on the Faddeev-Kulish \cite{Kulish1970} approach for the definition of soft-photon-dressed asymptotic states, following earlier work in \cite{Bloch1937,Kinoshita1962,Lee1964,Dollard1964,Chung1965,Kibble1968} (see also \cite{Jauch1955,Gaharia:2019xlh}).

The main idea is the definition of the interaction-picture asymptotic potential
\begin{eqnarray}\label{eq:aspot}
&&V^I_{as\,;\,QED}(t)\equiv\lim\limits_{|t|\rightarrow\pm\infty}\,V^I_{QED}(t)\,,
\end{eqnarray}
where $V^I_{QED}(t)$ is the linear interaction  term of the gauge field $A^\mu$ with the charged current:
\begin{eqnarray}\label{eq:intpotQED}
V^I_{QED}(t)=-\int d^{3}x \left[j^{\mu} A_{\mu}\right]\,.
\end{eqnarray}
The label $I$ on $V^I_{QED}$ means that we should substitute the \textit{interaction picture} mode expansions for the relevant fields (c.f. \eqref{eq:ABmodeq}). The dressed quantum states of QED are then given by
\begin{eqnarray}\label{eq:dressedE}
\left|p_1,\ldots,p_f\right\rrangle_{QED}&=&\mathcal{U}_{QED}\left|p_1,\ldots,p_f\right\rangle\nonumber\\[5pt]
\mathcal{U}_{QED}&\equiv&\mathcal{T}\,\exp{\left[-i\int_{0}^\infty\,dt\,\left(V^I_{as;QED}\right)\right]}\,.
\end{eqnarray}
The finite $S$-matrix for QED is given by
\begin{eqnarray}\label{eq:dressedS}
S^{finite}_{(1,\ldots, g|1,\ldots,f)}\equiv\left\llangle p_1,\ldots,p_g\right|S_{QED}\left|p_1,\ldots,p_f\right\rrangle\,.
\end{eqnarray}
where
\begin{eqnarray}\label{eq:SDQED}
S_{QED}=\mathcal{T}\,\exp{\left[-i\int_{-\infty}^\infty\,dt\,\left(V_{QED}\right)\right]}\,.
\end{eqnarray}
Is the usual Dyson S-matrix for QED. In short, we can calculate QED processes with the usual Feynman rules for QED derived from $S_{QED}$, as long as we use \textit{dressed} states as our asymptotic states. The result is guaranteed to be IR finite, with some subtleties that were recently addressed more carefully \cite{Hirai2019,Frye2019,Hannesdottir:2019opa,Hirai2021}. In this paper we will be rather cavalier with regards to these subtleties (e.g. using BRST instead of the Gupta-Bleuler condition), as they are not critical for the derivation of the Lorentz transformation rule for dressed charge-monopole states.

\subsection{Quantum Electro-Magneto Dynamics}
In this paper, we compute the IR dressing factors for QEMD. There are many formulations of this theory---by Schwinger \cite{Schwinger:1966nj}, Yan \cite{Yan1966}, Zwanziger \cite{Zwanziger:1968rs,Zwanziger:1970hk}, and Blagojevic and Sjevanovic \cite{Blagojevic1979}, to name just a few. They were all shown to be equivalent---for example in \cite{Blagojevic1988}. Here and below we use the local two-potential Lagrangian formulation due to Zwanziger \cite{Zwanziger:1968rs,Zwanziger:1970hk}. 

The Lagrangian for this theory is given in terms of the redundant vector fields $A^\mu$ and $B^\mu$, to which the electric (magnetic) current $j_e\,(j_g)$ couples as
\begin{eqnarray}
\mathcal{L}^I_{int}~=~-\left[j^\mu_eA_\mu+j^\mu_gB_\mu\right]\,.
\end{eqnarray}
Here the ''interaction picture" label $I$ on $\mathcal{L}^I_{int}$ means that we should substitute in it the mode expansions for the fields in the \textit{interaction picture}.
Though they seem to be separate degrees of freedom, the fields $A^\mu$ and $B^\mu$ are constrained and related. This is most explicitly expressed in their \textit{interaction picture} mode expansions in terms of creation and annihilation operators:
\begin{eqnarray}\label{eq:ABmodeq}
A_{\mu}(x)&=&\sum_{\lambda=\pm} \int \frac{d^{3} k}{(2 \pi)^{3}} \frac{1}{2 \omega_k}\left[\varepsilon_{\mu}^{* \lambda}(\vec{k}) a_{\lambda}(\vec{k}) e^{i k \cdot x}+\varepsilon_{\mu}^{\lambda}(\vec{k}) a_{\lambda}^{\dagger}(\vec{k}) e^{-i k \cdot x}\right]\nonumber\\[5pt]
B_{\mu}(x)&=&\sum_{\lambda=\pm} \int \frac{d^{3} k}{(2 \pi)^{3}} \frac{1}{2 \omega_k}\left[\widetilde{\varepsilon}_{\mu}^{* \lambda}(\vec{k}) a_{\lambda}(\vec{k}) e^{i k \cdot x}+\widetilde{\varepsilon}_{\mu}^{\lambda}(\vec{k}) a_{\lambda}^{\dagger}(\vec{k})e^{-i k \cdot x}\right]
\end{eqnarray}
where
\begin{eqnarray}
\left[a_{\lambda}(\vec{k}), a^{\dagger}_{\lambda'}\left(\vec{k}^{\prime}\right)\right]=\delta_{\lambda \lambda'}(2 \pi)^{3}\left(2 \omega_k\right) \delta^{3}\left(\vec{k}-\vec{k}^{\prime}\right)\,.
\end{eqnarray}
Here $\varepsilon_{\mu}^{\lambda}$ are the polarization vectors, $\varepsilon_{\lambda}^{\nu} \varepsilon_{\lambda' \nu}^{*}=\delta_{\lambda \lambda'}$, while $\widetilde{\varepsilon}_{\mu}^{a}$ are the dual polarization vectors satisfying
\begin{eqnarray}
k_{[\mu} \widetilde{\varepsilon}_{\nu]}^{\,\lambda *}=\epsilon_{\mu\nu}\left(k,\varepsilon_{\sigma}^{\lambda *}\right)\,.
\end{eqnarray}
These can be taken to be \cite{Colwell:2015wna,Strominger2016,Choi2020}
\begin{eqnarray}\label{eq:edual}
\widetilde{\varepsilon}_{\mu}^{~\lambda}=-A_{\mu \nu} \varepsilon^{\nu\,\lambda}, \quad A_{\mu \nu} \equiv  \frac{\epsilon_{\mu \nu}(n,k)}{n \cdot k+i\epsilon}\,,
\end{eqnarray}
where $n^\mu$ is an arbitrary spacelike 4-vector which stands for the direction of the Dirac string. We see that indeed $A^\mu$ and $B^\mu$ are not separate fields but rather different linear combinations of the same creation and annihilation operators for the photon. The relation \eqref{eq:edual} also shows that a gauge transformation which shifts $\varepsilon$ also shifts $\widetilde{\varepsilon}$, and so the gauge freedom is indeed just a $U(1)$ rather than two separate ones for $A^\mu$ and $B^\mu$.
For future reference, we also define
\begin{eqnarray}
a_\mu\equiv\sum_{\lambda=\pm}\,\varepsilon^{*\lambda}_\mu(\vec{k})a_\lambda(\vec{k})~~~,~~~
\widetilde{a}_\mu\equiv\sum_{\lambda=\pm}\,\widetilde{\varepsilon}^{*\lambda}_\mu(\vec{k})a_\lambda(\vec{k})\,,
\end{eqnarray}
as well as their hermitian conjugates. Finally, we can express the electromagnetic field strength and its dual as \cite{Zwanziger:1970hk}
\begin{eqnarray}\label{eq:Fs}
F^{\mu\nu}=\frac{1}{n^{2}}\left[F^{\mu\nu}_A-\widetilde{F}^{\mu\nu}_B\right]~~~,~~~
\widetilde{F}^{\mu\nu}=\frac{1}{n^{2}}\left[\widetilde{F}^{\mu\nu}_A+F^{\mu\nu}_B\right]\,,
\end{eqnarray}
where
\begin{eqnarray}\label{eq:FAs}
F^{\mu\nu}_A=n^{[\mu|}n^\rho(\partial_\rho A^{|\nu]}-\partial^{|\nu]} A_\rho)~~~,~~~
F^{\mu\nu}_B=n^{[\mu|}n^\rho(\partial_\rho B^{|\nu]}-\partial^{|\nu]} B_\rho)\,.
\end{eqnarray}
Substituting the mode expansions \eqref{eq:ABmodeq}, we have
\begin{eqnarray}\label{eq:FAsmodeq}
F^{\mu\nu}_A&=&in^{[\mu|}n^\rho\left\{\int \frac{d^{3} k}{(2 \pi)^{3}} \frac{1}{2 \omega_k}\left[(k_\rho a^{|\nu]}-k^{|\nu]} a_{\rho}) e^{i k \cdot x}-(k_\rho a^{\nu]\dagger}-k^{\nu]} a^\dagger_{\rho}) e^{-i k \cdot x}\right]\right\}\nonumber\\[5pt]
F^{\mu\nu}_B&=&in^{[\mu|}n^\rho\left\{ \int \frac{d^{3} k}{(2 \pi)^{3}} \frac{1}{2 \omega_k}\left[(k_\rho \widetilde{a}^{|\nu]}-k^{|\nu]} \widetilde{a}_{\rho}) e^{i k \cdot x}-(k_\rho \widetilde{a}^{\nu]\dagger}-k^{\nu]} \widetilde{a}^\dagger_{\rho}) e^{-i k \cdot x}\right]\right\}\,,
\end{eqnarray}
and so
\begin{eqnarray}\label{eq:Fmodeq}
F^{\mu\nu}&=&\frac{in^\rho}{n^{2}}\left\{\int \frac{d^{3} k}{(2 \pi)^{3}} \frac{1}{2 \omega_k}\left[(k_\rho n^{[\mu}a^{\nu]}-k_\rho\epsilon^{\mu\nu}(n,\widetilde{a})-n^{[\mu}k^{\nu]}a_{\rho}+\epsilon^{\mu\nu}(n,k)\widetilde{a}_{\rho}) e^{i k \cdot x}\right.\right.\nonumber\\[5pt]
&&~~~~~~~~~~~~~~~~~~~~~~~\left.\left.-(k_\rho n^{[\mu}a^{\nu]\dagger}-k_\rho\epsilon^{\mu\nu}(n,\widetilde{a}^\dagger)-n^{[\mu}k^{\nu]}a^\dagger_{\rho}+\epsilon^{\mu\nu}(n,k) \widetilde{a}^\dagger_{\rho}) e^{-i k \cdot x}\right]\right\}\nonumber\\[5pt]
\tilde{F}^{\mu\nu}&=&\frac{in^\rho}{n^{2}}\left\{\int \frac{d^{3} k}{(2 \pi)^{3}} \frac{1}{2 \omega_k}\left[(k_\rho \epsilon^{\mu\nu}(n,a)+k_\rho n^{[\mu}a^{\nu]}-\epsilon^{\mu\nu}(n,k)a_{\rho}-n^{[\mu}k^{\nu]}\widetilde{a}_{\rho}) e^{i k \cdot x}\right.\right.\nonumber\\[5pt]
&&~~~~~~~~~~~~~~~~~~~~~~~\left.\left.-(k_\rho\epsilon^{\mu\nu}(n,a^{\dagger})+k_\rho n^{[\mu}\widetilde{a}^{\nu]\dagger}-\epsilon^{\mu\nu}(n,ka^\dagger_{\rho}-n^{[\mu}k^{\nu]}\widetilde{a}^\dagger_{\rho}) e^{-i k \cdot x}\right]\right\}\,.\nonumber\\[5pt]
\end{eqnarray}

Similar interaction picture mode expansions exist for electrically and magnetically charged matter---though matter fields with different charges are indeed separate degrees of freedom and not linear combinations of the same creation and annihilation operators. For concreteness we consider here $f$ scalar fields $\phi_l(x)$ with $l=1,\ldots,f$ with electric and magnetic charges $(e_l,g_l)$. We then have the mode expansion
\begin{eqnarray}\label{eq:scalmodeq}
\phi_l(x)&=&\int \frac{d^{3} p}{(2 \pi)^{3}} \frac{1}{2 \omega_p}\left[b_l(\vec{p}) e^{i p \cdot x}+d_l^{\dagger}(\vec{p}) e^{-i p \cdot x}\right]\,,
\end{eqnarray}
where
\begin{eqnarray}
\left[b_l(\vec{p}), b_m^{\dagger}\left(\vec{p}^{\prime}\right)\right]&=&\left[d_l(\vec{p}), d_m^{\dagger}\left(\vec{p}^{\prime}\right)\right]=\delta_{lm}(2 \pi)^{3}\left(2 \omega_p\right) \delta^{3}\left(\vec{p}-\vec{p}^{\prime}\right)\,.
\end{eqnarray}

\subsection{Dressed States in QEMD}

After writing down the interaction potential and mode expansions for QEMD (in Zwanziger's two-potential language), we are now ready to compute the Faddeev-Kulish dressing factors in this theory. Previous authors \cite{Blagojevic1981,Blagojevic1982,Choi2020} have already calculated the IR dressing for QEMD in the one-potential formulation of \cite{Blagojevic1979}. However, these authors focused on the real dressing factor $R_{FK}$ and the way that it solves the IR problem for QEMD in a similar way to QED. Here we focus instead on the Faddeev-Kulish phase $\Phi_{FK}$ for QEMD. We show, that, in fact, this phase is associated with the extra electromagnetic angular momentum of the charge monopole system, and is directly linked to the pairwise little group.

To generalize the Faddeev-Kulish dressing factors to the electric-magnetic case, we start with the interaction-picture asymptotic interaction term in two-potential QEMD
\begin{eqnarray}\label{eq:intpot}
V^I_{a s\, ;\, QEMD}(t)=-\lim_{|t|\rightarrow\infty}\int d^{3}x \left[j_e^{\mu} A_{\mu} + j_g^{\mu} B_{\mu}\right]\,,
\end{eqnarray}
where 
\begin{eqnarray}\label{eq:currents}
j^{\mu}_e=i \sum_l\,e_l\left(\phi_l \partial^{\mu} \phi^{*}_l-\phi^{*}_l \partial^{\mu}\phi_l\right)~~~,~~~
j^{\mu}_g=i \sum_l\,g_l\left(\phi_l \partial^{\mu} \phi_l^{*}-\phi_l^{*} \partial^{\mu} \phi_l\right)\,.
\end{eqnarray}
Note that the label $I$ on $V^I_{a s\, ;\, QEMD}(t)$ means that we should substitute into \eqref{eq:intpot} the \textit{interaction picture} mode expansions \eqref{eq:ABmodeq} and \eqref{eq:scalmodeq}. A similar interaction term was also considered in \cite{Strominger2016}, in the context of the leading soft photon theorem for QEMD. Note also that we purposely leave out the ''seagull" interactions $A^\mu A_\mu\phi\phi^*$ and $B^\mu B_\mu\phi\phi^*$, since it is a non-generic feature of our choice of \textit{scalar} QEMD and gives an $\mathcal{O}(e^2g^2)$ contribution. We leave the study of this term for future work.

 Substituting the mode expansions for $\phi_l,\,A,\,B$, we get
\begin{eqnarray}\label{eq:mod}
&&V^I_{as\,;\,QEMD}(t)=-\sum_l\,\int D_l\,p\,\int \frac{d^{3} k}{(2 \pi)^{3}}\frac{1}{2 \omega_k} \nonumber\\[5pt]
&&\frac{p^{\mu}}{\omega_p}\left\{e_l\left[a_{\mu}(\vec{k}) e^{i \frac{p \cdot k}{\omega_p} t}+a_{\mu}^{\dagger}(\vec{k}) e^{-i \frac{p \cdot k}{\omega_p} t}\right]+g_l\left[\widetilde{a}_{\mu}(\vec{k}) e^{i \frac{p \cdot k}{\omega_p} t}+\widetilde{a}_{\mu}^{\dagger}(\vec{k}) e^{-i \frac{p\cdot k}{\omega_p} t}\right]\right\}\,,
\end{eqnarray}
where
\begin{eqnarray}
\rho_l(\vec{p})&\equiv&b_l(\vec{p})b_l^\dagger(\vec{p})-d_l^\dagger(\vec{p})d_l(\vec{p})\,,
\end{eqnarray}
are the scalar charge density operators, and
\begin{eqnarray}
\int D_l\,p\equiv\frac{d^{3} p}{(2\pi)^3}\frac{\rho_l(\vec{p})}{2\omega_p}\,.
\end{eqnarray}
The dressed electric-magnetic state in this case is
\begin{eqnarray}\label{eq:dressedEM}
\left|p_1,\ldots,p_f\right\rrangle&=&\mathcal{U}_{QEMD}\left|p_1,\ldots,p_f\right\rangle\,,
\end{eqnarray}
 with the dressing factor
\begin{eqnarray}\label{eq:UeqEM}
\mathcal{U}_{QEMD}&\equiv&\mathcal{T}\,\exp{\left[-i\int_{-\infty}^\infty\,dt\,V^I_{as\,;\,QEMD}(t)\right]}=e^{R_{FK}}\,e^{i\Phi_{FK}}\nonumber\\[5pt]
R_{FK}&=&-i\int_{-\infty}^\infty dt ~V^I_{as\,;\,QEMD}(t)\nonumber\\[5pt]  
\Phi_{FK}&=&\frac{i}{2} \int_{-\infty}^\infty dt_1\int_{-\infty}^{t_1}dt_2~[V^I_{as\,;\,QEMD}(t_1),V^I_{as\,;\,QEMD}(t_2)]\,.
\end{eqnarray}
Note that the lower limit $t=-\infty$ in the time integration of \eqref{eq:UeqEM}, which is different from the choice of $t=0$ in \cite{Hannesdottir:2019opa} (the lower limit is disregarded in the original Faddeev-Kulish paper \cite{Kulish1970}). Our choice of $t=-\infty$ reflects the fact that our dressed multiparticle states are defined independently of any S-matrix---they are simply a collection of plane waves for particles $1,\ldots,f$ embellished with all possible soft photons radiated from/between them. Our choice of lower limit is completely consistent with the BRST condition of \cite{Hirai2019}. It would be interesting to construct a full IR finite S-matrix given our choice of lower limit, and in particular to study its interplay with asymptotic symmetries in the spirit of \cite{Strominger2017,Hirai2021}.

\section{Dressed States as Pairwise States}\label{sec:dressedaspairwise}

In this section we derive one of the main results of this paper, namely that the dressed states defined in the previous section transform under Lorentz in the same way as the pairwise states from Section~\ref{sec:pairwisereview}. In other words, we want to show that for any Lorentz transformation $\Lambda$
\begin{eqnarray}\label{eq:toshow0}
U\left[\Lambda\right]\,\left|p_1,\ldots,p_f\right\rrangle=e^{i\Phi_{LG}}\left|\Lambda p_1,\ldots,\Lambda p_f\right\rrangle\,,
\end{eqnarray}
where $\Phi_{LG}\equiv\sum_{l<m}q_{lm}\varphi_{LG}(p_l,p_m,\Lambda)$ and $\varphi_{LG}$ is the pairwise little group phase defined in Section~\ref{sec:pairwisereview}. Note that the unitary representation $U\left[\Lambda\right]$ in the above equation is no longer our choice, but is actually uniquely defined by the action of our theory, two potential QEMD \cite{Zwanziger:1970hk}. In particular, for an infinitesimal Lorentz transformation $\Lambda^\mu_\nu=\exp(\delta\tau\omega^\mu_\nu)$, its unitary representation is given by  
\begin{eqnarray}\label{eq:ULa}
U\left[\Lambda\right]=\exp\left[\frac{i}{2}\,\delta\tau\,M^{\mu\nu}\omega_{\nu\mu}\right]\,,
\end{eqnarray}
where $M^{\mu\nu}$ is the Noether generator of Lorentz transformations in QEMD, whose explicit expression is given below. Substituting the infinitesimal transformation \eqref{eq:ULa} in \eqref{eq:toshow0} and presenting the dressing explicitly, we have
\begin{eqnarray}\label{eq:toshow01}
\exp\left[\frac{i}{2}\,\delta\tau\,M^{\mu\nu}\omega_{\nu\mu}\right]\,e^{R_{FK}}\,e^{i\Phi_{FK}}\,\left|p_1,\ldots,p_f\right\rangle=e^{i\Phi_{LG}}\,e^{R_{FK}}\,e^{i\Phi_{FK}}\,\left|\Lambda p_1,\ldots,\Lambda p_f\right\rangle\,.
\end{eqnarray}
As we shall see below, the phase factor $\Phi_{FK}$ evaluates to a c-number when acting on the multiparticle state to its right, and so we can commute it freely. Rearranging, we have
\begin{eqnarray}\label{eq:toshow1}
e^{-R_{FK}}\exp\left[\frac{i}{2}\,\delta\tau\,M^{\mu\nu}\omega_{\nu\mu}\right]\,e^{R_{FK}}\,e^{-i\Delta\Phi_{FK}}\,\left|p_1,\ldots,p_f\right\rangle\,=e^{i\Phi_{LG}}\left|\Lambda p_1,\ldots,\Lambda p_f\right\rangle\,,
\end{eqnarray}
where $\Delta\Phi_{FK}=\Phi_{FK}|_{\Lambda p}-\Phi_{FK}|_{p}$.
We proceed by expanding both sides to leading order in $\delta\tau$. For future reference we define
\begin{eqnarray}\label{eq:phiexpand}
\Delta \Phi_{FK}&\equiv&\frac{\delta\tau}{2}\,\omega_{\mu\nu}\,\Delta\Phi^{\mu\nu}_{FK}+\mathcal{O}(\delta\tau^2)\,.
\end{eqnarray}
We then have
\begin{eqnarray}\label{eq:toshow2}
\left\{e^{-R_{FK}}M^{\mu\nu}e^{R_{FK}}-\Delta\Phi^{\mu\nu}_{FK}\right\}\left|p_1,\ldots,p_f\right\rangle=\Phi^{\mu\nu}_{LG}\,\left|p_1,\ldots,p_f\right\rangle\,,
\end{eqnarray}
where $\Phi^{\mu\nu}_{LG}$ is given in \eqref{eq:PhimunuLG}.
We will now prove \eqref{eq:toshow2} by explicit calculation, using the explicit expressions for $R_{FK},\,\Delta\Phi_{FK}$ and $M^{\mu\nu}$ in QEMD. As a first step
we can use the Baker-Campbell-Hausdorff lemma on the left hand side,
\begin{eqnarray}\label{eq:toshow3}
\left\{\left[M^{\mu\nu},R_{FK}\right]+\frac{1}{2}\left[\left[M^{\mu\nu},R_{FK}\right],R_{FK}\right]-\Delta\Phi^{\mu\nu}_{FK}\right\}\left|p_1,\ldots,p_f\right\rangle=\Phi^{\mu\nu}_{LG}\,\left|p_1,\ldots,p_f\right\rangle\,.\nonumber\\
\end{eqnarray}
Note that higher commutators in the expansion vanish since the second commutator is already a c-number. We now turn to calculate the left hand side of this equation. We begin by evaluating $\Delta\varphi_{FK}$, and later recall the expression for $M^{\mu\nu}$ and calculate its commutators with $R_{FK}$.

\subsection{Calculating $\Delta\Phi^{\mu\nu}_{FK}$}
To calculate the shift $\Delta\Phi^{\mu\nu}_{FK}$ in the dressing phase following an infinitesimal Lorentz transformation, we first need an expression for $\Phi_{FK}$.
As a first step to calculating $\Phi_{FK}$, we bring \eqref{eq:UeqEM} to a form more fit for Feynman integration in the following way
\begin{eqnarray}\label{eq:PhiFeyn}
\Phi_{FK}&=&\frac{i}{4}\left\{ \int_{-\infty}^\infty dt_1\int_{-\infty}^{t_1}dt_2~[V_{as\,;\,QEMD}^{I}(t_1),V_{as\,;\,QEMD}^{I}(t_2)]\,+\right.\nonumber\\[5pt]
&&~~~~\left.\int_{-\infty}^\infty dt_2\int_{-\infty}^{t_2}dt_1~[V_{as\,;\,QEMD}^{I}(t_2),V_{as\,;\,QEMD}^{I}(t_1)]\right\}\nonumber\\[5pt]
&=&\frac{i}{4}\int_{-\infty}^\infty dt_1\int_{-\infty}^{\infty}dt_2~[V_{as\,;\,QEMD}^{I}(t_{max}),V_{as\,;\,QEMD}^{I}(t_{min})]\,,
\end{eqnarray}
where $t_{max}=max(t_1,t_2),\,t_{min}=min(t_1,t_2)$. To explicitly evaluate the commutator in \eqref{eq:PhiFeyn} we need to evaluate polarization sums coming from the mode expansions \eqref{eq:ABmodeq}. Specializing to quantum states that satisfy the free Gupta-Bleuler condition $k^\mu a_\mu\left|\psi\right\rangle=0$, we have
\begin{eqnarray}\label{eq:prop}
\varepsilon^{\alpha*}_\mu\varepsilon_{\alpha\nu}=g_{\mu\nu}~~~\rightarrow~~~\varepsilon^{\alpha*}_\mu\widetilde{\varepsilon}_{\alpha\nu}=-\frac{\epsilon_{\mu\nu}(n,k)}{n\cdot k+i\epsilon}\,,
\end{eqnarray}
this form of the magnetic propagator is unique up to gauge transformations which rotate $n$ and could also change the $\epsilon$ prescription on the spurious $n\cdot k$ pole. An alternative $\epsilon$ prescription, where $(n\cdot k+i\epsilon)^{-1}\rightarrow \tfrac{1}{2}\left[(n\cdot k+i\epsilon)^{-1}+(n\cdot k-i\epsilon)^{-1}\right]$, is equivalent to a two-sided Dirac string, and does not change the results derived in this paper. 

The expression \eqref{eq:PhiFeyn} then evaluates to 
\begin{eqnarray}\label{eq:Ct1}
\Phi_{FK}&=&4\pi\sum_{l<m}\, q_{lm}\,\iint D_l\,p_a\,D_m\,p_b\,\int_{-\infty}^\infty \frac{dt_1}{\omega_a}\int_{-\infty}^{\infty}\frac{dt_2}{\omega_b}\,\text{Im}\left[I(p_a,p_b,n)\right]\,\nonumber\\[5pt]
I(p_1,p_2,p_3)&\equiv&-\int\frac{d^{4} k}{(2 \pi)^{4}}\frac{i\epsilon(p_1,p_2,p_3, k)}{(k^2+i\epsilon)(p_3\cdot k+i\epsilon)}\,e^{-ik\cdot\Delta_{12}(p_a,p_b)}\,,
\end{eqnarray}
where $\Delta_{12}^\mu(a,b)=\frac{t_{1}a^\mu}{\omega_a}-\frac{t_{2}b^{\mu}}{\omega_b}$, $q_{lm}=(e_lg_m-e_mg_l)/4\pi$, and $p_3=n$. By a change of integration variables, it's easy to see that
\begin{eqnarray}\label{eq:QM2}
\Phi_{FK}&=&\sum_{l<m}\, q_{lm}\,\iint D_l\,p_a\,D_m\,p_b~\left[\varphi_{FK}(p_a,p_b,n)\right]\,.
\end{eqnarray}
where
\begin{eqnarray}\label{eq:II}
&&\varphi_{FK}(p_1,p_2,p_3)=4\pi\,\text{Im}\left[\mathcal{I}(p_1,p_2,p_3)-\mathcal{I}(-p_1,p_2,p_3)-\mathcal{I}(p_1,-p_2,p_3)+\mathcal{I}(-p_1,-p_2,p_3)\right]\nonumber\\[5pt]
&&\mathcal{I}(p_1,p_2,p_3)=\iint_0^\infty\frac{dt_1}{\omega_1}\frac{dt_2}{\omega_2}\,I(p_1,p_2,p_3)\nonumber\\[5pt]
&&~~~~~~~~~~~~~~~=\int\frac{d^{4} k}{(2 \pi)^{4}}\frac{i\epsilon(p_1,p_2,p_3, k)}{(k^2+i\epsilon)(p_1\cdot k-i\epsilon)(p_2\cdot k+i\epsilon)(p_3\cdot k+i\epsilon)}\,.
\end{eqnarray}
In complete agreement with the calculation \cite{Terning:2018udc} of soft-photon resummation using the Weinberg formalism \cite{Weinberg:1995mt}. Note that it is this integral that contains the topological linking number \cite{Terning:2018udc} of the pair.
As we shall see below, this integral is not well defined and needs regularization. However, here we are actually interested only in the variation of $\mathcal{I}$ under a Lorentz transformation $\Lambda$. This variation is well defined and does not require regularization. In fact, this variation gives us exactly the pairwise LG phase. To see this, we calculate
\begin{eqnarray}\label{eq:DmathI}
\Delta \varphi_{FK}&\equiv&\varphi_{FK}(\Lambda p_1,\Lambda p_2,p_3)-\varphi_{FK}(p_1,p_2,p_3)\,.
\end{eqnarray}
One can easily show that $\mathcal{I}(\Lambda p_1,\Lambda p_2,p_3)=\mathcal{I}(p_1,p_2,\Lambda^{-1}p_3)$ by a simple change of integration variables, namely
\begin{eqnarray}\label{eq:LI}
\mathcal{I}(\Lambda p_1,\Lambda p_2,p_3)&=&\int\frac{d^{4} k}{(2 \pi)^{4}}\,\frac{i\epsilon(p_1,p_2,\Lambda^{-1}p_3,\Lambda^{-1} k)}{\left[k^2+i\epsilon\right]\left[(\Lambda p_1)\cdot k-i\epsilon\right]\left[(\Lambda p_2)\cdot k+i\epsilon\right]\left[p_3\cdot k+i\epsilon\right]}\,.\nonumber\\[5pt]
&=&\int\frac{d^{4} k}{(2 \pi)^{4}}\,\frac{i\epsilon(p_1,p_2,\Lambda^{-1}p_3,k)}{\left[k^2+i\epsilon\right]\left[p_1\cdot k-i\epsilon\right]\left[p_2\cdot k+i\epsilon\right]\left[(\Lambda^{-1}p_3)\cdot k+i\epsilon\right]}\nonumber\\[5pt]
&=&I(p_1,p_2,\Lambda^{-1}p_3)\,.
\end{eqnarray}
Consequently, we have
\begin{eqnarray}\label{eq:DLI}
\Delta \mathcal{I}&=&-i\int\frac{d^{4} k}{(2 \pi)^{4}}\,\frac{T^{\mu\nu}(p_1,p_2,p_3,p_4)\,k_\mu k_\nu}{(k^2+i\epsilon)(p_1\cdot k-i\epsilon)(p_2\cdot k+i\epsilon)(p_3\cdot k+i\epsilon)(p_4\cdot k+i\epsilon)}\,,\nonumber\\
\end{eqnarray}
where $p_4=\Lambda^{-1}p_3$, and 
\begin{eqnarray}\label{eq:Dec}
T^{\mu\nu}(p_1,p_2,p_3,p_4)&=&\frac{1}{2}\left[{p}^{\{\mu}_4\epsilon^{\nu\}}(p_1,p_2,p_3)-p^{\{\mu}_3\epsilon^{\nu\}}(p_1,p_2,p_4)\right]\,,
\end{eqnarray}
is a symmetric tensor\footnote{Here and below, $a^{\{\mu}b^{\nu\}}=a^\mu b^\nu+a^\nu b^\mu$ and $a^{[\mu}b^{\nu]}=a^\mu b^\nu-a^\nu b^\mu$.}. We can decompose this tensor as
\begin{eqnarray}\label{eq:Dec2}
T^{\mu\nu}(p_1,p_2,p_3,p_4)&=&\bar{\epsilon}\,\eta^{\mu\nu}+i\sum_{i=1}^4\,f_{1i}\,p^{\{\mu}_1p^{\nu\}}_i+i\sum_{i=1}^4\,f_{2i}\,p^{\{\mu}_2p^{\nu\}}_i\,.
\end{eqnarray}
Here we define $\bar{\epsilon}\equiv \epsilon(p_1,p_2,p_3,p_4)$ and $\epsilon^\mu(p_1,...,\xcancel{p}_i,...,p_4)\equiv (-1)^{i}\epsilon^\mu(p_1,...,p_{i-1},p_{i+1},...,p_4)$, then the coefficients $f_{ij}$ are defined as
\begin{eqnarray}\label{eq:epBasis}
f_{ij}=i\frac{\epsilon^\mu(p_1,...,\xcancel{p}_i,...,p_4)\epsilon_\mu(p_1,...,\xcancel{p}_j,...,p_4)}{2\bar{\epsilon}}\,,
\end{eqnarray}
Consequently, we can write
\begin{eqnarray}\label{eq:DLI2}
\Delta \mathcal{I}&=&\Delta\mathcal{I}_*+f_{12} \Delta\mathcal{I}_{12}+f_{31} \Delta\mathcal{I}_{31}+f_{14}\Delta\mathcal{I}_{14}+f_{11}\Delta\mathcal{I}_{11}\nonumber\\[5pt]
&&~~~~~\,+f_{12}\Delta \mathcal{I}_{12}+f_{23}\Delta \mathcal{I}_{23}+f_{24}\Delta\mathcal{I}_{24}+f_{22}\Delta\mathcal{I}_{22}\,,
\end{eqnarray}
where
\begin{eqnarray}\label{eq:DLI2def}
\Delta \mathcal{I}_*
&=&\int\frac{d^{4} k}{(2 \pi)^{4}}\,\frac{-i\bar{\epsilon}}{(p_1\cdot k-i\epsilon)(p_2\cdot k+i\epsilon)(p_3\cdot k+i\epsilon)(p_4\cdot k+i\epsilon)}\nonumber\\[5pt]
\Delta \mathcal{I}_{ij}
&=&\int\frac{d^{4} k}{(2 \pi)^{4}}\,\frac{(p_i\cdot k)(p_j\cdot k)}{(k^2+i\epsilon)(p_1\cdot k-i\epsilon)(p_2\cdot k+i\epsilon)(p_3\cdot k+i\epsilon)(p_4\cdot k+i\epsilon)}\,.
\end{eqnarray}
We can explicitly check that all of the $\mathcal{I}_{ij}$ sum up to zero, and only $\Delta\mathcal{I}_*$ remains. To see this, we define the master integral
\begin{eqnarray}\label{eq:lowmaster}
I(v_1,v_2)&=&\int\frac{d^{4} k}{(2 \pi)^{4}}\,\frac{1}{(k^2+i\epsilon)(v_1\cdot k+i\epsilon)(v_2\cdot k+i\epsilon)}\,.
\end{eqnarray}
In terms of this master integral, we have (see appendix~\ref{app:red} for the detailed calculation)
\begin{eqnarray}\label{eq:lowmaster1}
&&\Delta \mathcal{I}_{11}=-\frac{f_{31}I(p_2,p_4)+f_{12}I(p_3,p_4)+f_{14}I(p_2,p_3)}{f_{11}}\nonumber\\[5pt]
&&\Delta \mathcal{I}_{22}=-\frac{f_{12}I(p_3,p_4)-f_{24}I(-p_1,p_3)-f_{23}I(-p_1,p_4)}{f_{22}}\nonumber\\[5pt]
&&\Delta \mathcal{I}_{12}=I(p_3,p_4)\nonumber\\[5pt]
&&\Delta \mathcal{I}_{13}=I(p_2,p_4)~~~~,~~~~\Delta \mathcal{I}_{23}=-I(-p_1,p_4)\nonumber\\[5pt]
&&\Delta \mathcal{I}_{14}=I(p_2,p_3)~~~~,~~~~\Delta \mathcal{I}_{24}=-I(-p_1,p_3)\,.
\end{eqnarray}
Substituting this expansion back in \eqref{eq:DLI2}, we see that all of the $\mathcal{I}_{ij}$ cancel out and so by \eqref{eq:II},\eqref{eq:DmathI}, and \eqref{eq:DLI2},
\begin{eqnarray}\label{eq:DLI3}
&&\Delta \varphi_{FK}=-4\pi\,\text{Im}\left[~~\Delta \mathcal{I}_*(p_1,p_2,p_3)-\Delta \mathcal{I}_*(-p_1,p_2,p_3)\right.\nonumber\\[5pt]
&&~~~~~~~~~~~~~~~~~~~~\left.-\Delta \mathcal{I}_*(p_1,-p_2,p_3)+\Delta \mathcal{I}_*(-p_1,-p_2,p_3)\right]\,,
\end{eqnarray}
or in other words,
\begin{eqnarray}\label{eq:DLI4}
\Delta \varphi_{FK}&=&4\pi\,\text{Im}\left[\Delta\mathcal{I}_{**}\right]\nonumber\\[5pt]
\Delta\mathcal{I}_{**}&\equiv&-i\bar{\epsilon}\int\frac{d^4k}{(2\pi)^2}\,\frac{\delta\left(p_1\cdot k\right)\,\delta\left(p_2\cdot k\right)}{(p_3\cdot k+i\epsilon)(p_4\cdot k+i\epsilon)}\,,
\end{eqnarray}
where we have use the ``cutting identity"
\begin{eqnarray}\label{eq:cutid}
\delta(a)=2\pi i \,\left(\frac{1}{a-i\epsilon}-\frac{1}{a+i\epsilon}\right)\,.
\end{eqnarray}
This integral is calculated explicitly in appendix~\ref{app:miscIstst}, with the result
\begin{eqnarray}\label{eq:DLI4result}
\Delta\mathcal{I}_{**}&=&\frac{i}{2\pi}\arccos\left[\hat{\epsilon}(p_1,p_2,p_3)\cdot \hat{\epsilon}(p_1,p_2,p_4)\right]\,.
\end{eqnarray}
Note that while Eq. (\ref{eq:lowmaster}) was naively log divergent, $\Delta\mathcal{I}_{**}$ is finite, a sign that the final answer is not sensitive to UV photons, as expected.
The resulting shift in the Faddeev-Kulish phase is then
\begin{eqnarray}\label{eq:QM25}
\Delta\varphi_{FK}(p_1,p_2,n)&=&2\arccos\left[\hat{\epsilon}(p_1,p_2,\Lambda^{-1}n)\cdot\hat{\epsilon}(p_1,p_2, n)\right]\,.
\end{eqnarray}
But this is exactly twice the little group phase $\varphi_{LG}$ in \eqref{eq:pLGphase}. Consequently, we have
\begin{eqnarray}\label{eq:QM26}
\Delta\Phi^{\mu\nu}_{FK}&=&\sum_{l<m}q_{lm}\,\Delta \varphi^{\mu\nu}_{FK;lm}=2\sum_{l<m}q_{lm}\,\varphi^{\mu\nu}_{LG;lm}=-2\Phi^{\mu\nu}_{LG}\,,
\end{eqnarray}
where the minus sign is a consequence of \eqref{eq:PhiLG}. This already gives a part of the required contribution on the left hand side of \eqref{eq:toshow3}. The other half comes from the commutators of the Lorentz generator $M^{\mu\nu}$, which we now define.
\subsection{The Angular Momentum Operator in Two-Potential QEMD}
The energy momentum tensor in two-potential QEMD is given by
\cite{Zwanziger:1970hk}:
\begin{eqnarray}\label{eq:thetas}
\theta^{\mu \nu}&=&\theta^{\mu\nu}_{EM}+\theta^{\mu\nu}_{\varphi,A}+\theta^{\mu\nu}_{\varphi,B}+\theta^{\mu\nu}_{g.f.}  
-n^{\mu}\epsilon^{\nu}\left(n,\left(n\cdot \partial\right)^{-1}j_e,\left(n\cdot \partial\right)^{-1}j_g\right)\nonumber\\[5pt]
\theta^{\mu\nu}_{EM}&=&\frac{1}{2}\left(F^{\mu}_{~\alpha} F^{\alpha\nu}+\widetilde{F}^{\mu}_{~\alpha} \widetilde{F}^{\alpha\nu}\right)\nonumber\\[5pt]
\theta^{\mu\nu}_{\varphi,V}&=&\sum_l\,\frac{1}{2}\left(D^{\{\mu}_{V,l}\phi_l\right)\left(D^{\nu\}}_{V,l}\phi_l\right)^*-\frac{1}{2}\eta^{\mu\nu}\left[\eta_{\alpha\beta}\left(D^{\alpha}_{V,l}\phi_l\right)\left(D^\beta_{V,l}\phi_l\right)^*-m^2_{l}\phi_l\phi_l^*\right]\,,
\end{eqnarray}
where $V=A,B$. $D^\mu_{A,l}=\partial^\mu-ie_lA^\mu$ and $D^\mu_{B,l}=\partial^\mu-ig_lB^\mu$ are the electric and magnetic covariant derivatives. In the quantum theory, $\theta^{\mu\nu}$ is promoted to an operator acting on multiparticle quantum states. We will not need an explicit expression for the gauge fixing term $\theta^{\mu\nu}_{g.f.}$, and we refer the reader to \cite{Zwanziger:1970hk} for its explicit form. We can choose our physical Hilbert space such that the last term vanishes as an operator on all physical states. In this case $\theta^{\mu \nu}$ is a symmetric operator and implies conserved angular momentum (Lorentz) operators. Specifically, we have
\begin{eqnarray}\label{eq:totang}
&&M^{\mu\nu}\equiv M^{\mu\nu}_{EM}+M^{\mu\nu}_{matter}+M^{\mu\nu}_{g.f.}~~~,~~M^{\mu\nu}_{i}=\int d^3x\,  x^{[\mu} \theta^{0\nu]}_{i}\,.
\end{eqnarray}
For later reference, we now insert the mode expansions and \eqref{eq:scalmodeq} into \eqref{eq:totang} to get a mode expansion of $M^{\mu\nu}_{matter}$. We get
\begin{eqnarray}\label{eq:matangk}
M^{\mu\nu}_{matter}&=&M^{\mu\nu}_{kin}+M^{\mu\nu}_{A}+M^{\mu\nu}_{A^2}+(A,e_l\leftrightarrow B,g_l)\nonumber\\[5pt]
M^{\mu\nu}_{A}&=&i \sum_l\,e_l\,\int D_l\,p\,\int \frac{d^3k}{(2\pi)^3}\frac{1}{2\omega_k}\,\frac{tp^{[\mu}}{\omega_p}\left\{a^{\nu]}(k)e^{ik\cdot \frac{p t}{\omega_p}}+a^{\nu]\dagger}(k)e^{-ik\cdot \frac{p t}{\omega_p}}\right\}\,\nonumber\\[5pt]
M^{\mu\nu}_{B}&=&i \sum_l\,g_l\,\int D_l\,p\,\int \frac{d^3k}{(2\pi)^3}\frac{1}{2\omega_k}\,\frac{tp^{[\mu}}{\omega_p}\left\{\widetilde{a}^{\nu]}(k)e^{ik\cdot \frac{p t}{\omega_p}}+\widetilde{a}^{\nu]\dagger}(k)e^{-ik\cdot \frac{p t}{\omega_p}}\right\}\,.\nonumber\\
\end{eqnarray}
We remind the reader that $D_l\,p\equiv \frac{d^3p}{(2\pi)^3}\frac{\rho_l(p)}{2\omega_p}$. We will not be interested in $M^{\mu\nu}_{kin}$, as it only reflects the orbital part of the angular momentum and will not have any nontrivial commutator with $R_{FK}$. Similarly, we drop the angular momentum coming from the second term in the expression for $\theta^{\mu\nu}_{\varphi_V}$ in \eqref{eq:thetas}. When we commute $M^{\mu\nu}_A$ with $R_{FK}$, this term gives a vanishing contribution at $\mathcal{O}(eg)$, and so we drop it here for simplicity. Finally, we also do not explicitly consider $M^{\mu\nu}_{A^2}$ here, since it is a non-generic feature of our choice of \textit{scalar} QEMD and gives an $\mathcal{O}(e^2g^2)$ contribution. We leave the study of this term for future work.

\subsection{Angular Momentum Commutators}
In this section we prove \eqref{eq:toshow2}, which demonstrates that dressed and pairwise states transform the same way. To this end, we explicitly evaluate 
\beq
[M^{\mu\nu}_A+M^{\mu\nu}_B,R_{FK}] \quad  {\rm and} \quad  \tfrac{1}{2}[[M^{\mu\nu}_{EM},R_{FK}],R_{FK}]~.
\eeq
One can explicitly check that these are the only terms that contribute to $\Delta \varphi^{\mu\nu}_{lm}$ in \eqref{eq:toshow3}. Since the former commutators essentially evaluate the angular momentum associated with the retarded fields sourced by the charge and the monopole, the final parts of our derivation coincide with sections 2 and 4 of \cite{Zwanziger:1972sx}.

\subsubsection{$[M^{\mu\nu}_A+M^{\mu\nu}_B,R_{FK}]$}\label{sec:IA}
Since $M^{\mu\nu}_A\sim e_l$, we focus here on the $\mathcal{O}(g_m)$ contribution to the retarded potential. This is the term responsible for the extra angular momentum in the electromagnetic field associated with the charge-monopole pair. We have
\begin{eqnarray}
\left[M^{\mu\nu}_A,R_{FK}\right]&=&
\sum_{l<m}\,q_{lm}\,\iint\,D_l\,p_a\,D_m\,p_b~\left[\mathcal{I}_{A}(p_a,p_b,n)\right]\nonumber\\[5pt]
\mathcal{I}_{A}(p_1,p_2,p_3)&=&-\frac{8\pi t}{\omega_1}\,\int_{-\infty}^\infty\,\frac{dt'}{\omega_2}\,\,\text{Im}\left\{\int \frac{d^{3} k}{(2 \pi)^{3}} \frac{1}{2 \omega_k}\,\frac{p^{[\mu}_1\epsilon^{\nu]}(p_2,p_3,k)}{p_3\cdot k +i\epsilon}e^{-ik\cdot\left(\frac{p^\mu_1t}{\omega_1}-\frac{p^\mu_2t'}{\omega_2}\right)}\right\}\,,\nonumber\\
\end{eqnarray}
where $\mathcal{I}_A$ is evaluated in appendix~\ref{app:miscIA}, with the result that
\begin{eqnarray}
\mathcal{I}_{A}(p_1,p_2,p_3)&=&\frac{p^{[\mu}_1\epsilon^{\nu]}\left(p_1,p_2,p_3\right)}{\tau_{12}}\,\frac{m^2_2\,p_{31}-p_{12}p_{23}}{\epsilon^2(p_1,p_2,p_3)}\,.
\end{eqnarray}
Similarly, the $\mathcal{I}_B$ contribution from $B^\mu$ is given by $-\mathcal{I}_A|_{1\leftrightarrow 2}$.
Together, these two contributions combine into \cite{Zwanziger:1972sx}
\begin{eqnarray}\label{eq:IAB}
\mathcal{I}_{A}+\mathcal{I}_{B}&=&
\frac{n^{[\mu}\epsilon^{\nu]}\left(p_1,p_2,p_3\right)}{\epsilon^2(p_1,p_2,p_3)}-\frac{\epsilon^{\mu\nu}(p_1,p_2)}{\tau_{12}}\,.
\end{eqnarray}
And so the relevant piece of the angular momentum commutator is
\begin{eqnarray}\label{eq:MAB}
&&\left[M^{\mu\nu}_A+M^{\mu\nu}_B,R_{FK}\right]=\sum_{l<m}\,q_{lm}\,\iint\,D_l\,p_a\,D_m\,p_b~\left\{\frac{n^{[\mu}\epsilon^{\nu]}\left(p_a,p_b,n\right)}{\epsilon^2(p_a,p_b,n)}-\frac{\epsilon^{\mu\nu}(p_a,p_b)}{\tau_{ab}}\right\}\,.\nonumber\\
\end{eqnarray}
\subsubsection{$\tfrac{1}{2}\left[\left[M^{\mu\nu}_{EM},R_{FK}\right],R_{FK}\right]$}\label{sec:IF}
This commutator is given by
\begin{eqnarray}\label{eq:EMang}
&&\tfrac{1}{2}\left[\left[M^{\mu\nu}_{EM},R_{FK}\right],R_{FK}\right]=\frac{1}{2}\int d^3x\,x^{[\mu}\left\{\left[F^{0}_{~\alpha},R_{FK}\right]\left[F^{\alpha|\nu]},R_{FK}\right]+\left(F\leftrightarrow\widetilde{F}\right)\right\}\,\nonumber\\[5pt]
\end{eqnarray}
Note that this is the only nonzero contribution, since $M^{\mu\nu}_{EM}$ is bilinear in EM creation/annihilation operators. The commutators $\left[F^{\mu\nu},R_{FK}\right]$ $\left(\left[\widetilde{F}^{\mu\nu},R_{FK}\right]\right)$ has a clear physical meaning---it is the retarded field strength (dual field strength) generated by the asymptotic particles in the quantum state. Substituting the mode expansion \eqref{eq:Fmodeq} and the definition of $R_{FK}$ in \eqref{eq:UeqEM} and \eqref{eq:mod}, we have
\begin{eqnarray}\label{eq:EMangcom1}
\left[F^{\mu\nu}_{EM},R_{FK}\right]&=&\sum_l\,\int\,D_l\,p~\left[\mathcal{I}_{F,l}(p)\right]\nonumber\\[5pt]
\mathcal{I}_{F,l}(p)&\equiv&-2i\int_{-\infty}^\infty dt'~\text{Re}\left\{\int \frac{d^{3} k}{(2 \pi)^{3}} \frac{1}{2 \omega_k}~\left[e_lk^{[\mu}p^{\nu]}-g_l\epsilon^{\mu\nu}\left(k,p\right) \right]e^{-ik\cdot(x-\frac{p}{\omega_p} t')}\right\}\,.\nonumber\\[5pt]
\end{eqnarray}
Note that the $n^\mu$ dependence has dropped off, as it should for all gauge invariants like the retarded electromagnetic field strength. The factor of $1/2$ here comes from the one in our general formula for the dressing, \eqref{eq:UeqEM}. We calculate this integral explicitly in appendix~\ref{app:miscIFl}, and get the retarded field strength 
\begin{eqnarray}\label{eq:EMangcom2txt}
\mathcal{I}_{F,l}(p)=-\frac{m^2}{4\pi}\,\frac{e_l\,x^{[\mu}p^{\nu]}-g_l\,\epsilon^{\mu\nu}\left(x,p\right)}{{\left[(p\cdot x)^2-m^2x^2\right]}^{\frac{3}{2}}}\,,
\end{eqnarray}
where $m=p^2$. Similarly,
\begin{eqnarray}\label{eq:EMangcom3}
\mathcal{I}_{\widetilde{F},l}(p)&=&-\frac{m^2}{4\pi}\,\frac{e_l\,\epsilon^{\mu\nu}\left(x,p\right)+g_l\,x^{[\mu}p^{\nu]}}{{\left[(p\cdot x)^2-m^2x^2\right]}^{\frac{3}{2}}}\,.
\end{eqnarray}
Substituting in \eqref{eq:EMang} and keeping only $\mathcal{O}(e_lg_m)$ terms (the $\mathcal{O}(e^2_l,g^2_l)$ terms vanish by symmetry), we have
\begin{eqnarray}\label{eq:EMangs}
\left[M^{\mu\nu}_{EM},R_{FK}\right]&=&\sum_{l<m}\,q_{lm}\,\iint\,D_l\,p_a\,D_m\,p_b~\left[\mathcal{I}^{\mu\nu}_{EM}(p_a,p_b)\right]\nonumber\\[5pt]
\mathcal{I}^{\mu\nu}_{EM}(p_a,p_b)&\equiv&\frac{m^2_am^2_bt}{4\pi}\int d^3x\,\frac{x^{[\mu}\epsilon^{\nu]}(x,p_a,p_b)}{{\left[(p_a\cdot x)^2-m^2_ax^2\right]}^{\frac{3}{2}}{\left[(p_b\cdot x)^2-m^2_bx^2\right]}^{\frac{3}{2}}}\,.
\end{eqnarray}
The integral $\mathcal{I}^{\mu\nu}_{EM}$ is also calculated explicitly in appendix~\ref{app:miscIEM}. The result is 
\begin{eqnarray}\label{eq:EMangsftxt}
\mathcal{I}^{\mu\nu}_{EM}&=&\frac{\epsilon^{\mu\nu}(p_a,p_b)}{\tau_{ab}}\,,
\end{eqnarray}
and so we have
\begin{eqnarray}\label{eq:EMangs5}
\left[\left[M^{\mu\nu}_{EM},R_{FK}\right],R_{FK}\right]&=&\sum_{l<m}\,q_{lm}\,\iint\,D_l\,p_a\,D_m\,p_b~\left[\frac{\epsilon^{\mu\nu}(p_a,p_b)}{\tau_{ab}}\right] \,.
\end{eqnarray}

Summing up the contributions \eqref{eq:MAB} and \eqref{eq:EMangs5}, we arrive at \begin{eqnarray}\label{eq:Mtot}
&&\left[M^{\mu\nu},R_{FK}\right]+\frac{1}{2}\left[\left[M^{\mu\nu},R_{FK}\right],R_{FK}\right]=\sum_{l<m}\,q_{lm}\,\iint\,D_l\,p_a\,D_m\,p_b~\left[\frac{n^{[\mu}\epsilon^{\nu]}\left(p_a,p_b,n\right)}{\epsilon^2(p_a,p_b,n)}\right]\,,\nonumber\\
\end{eqnarray}
and so 
\begin{eqnarray}\label{eq:Mtotres}
&&\left\{\left[M^{\mu\nu},R_{FK}\right]+\frac{1}{2}\left[\left[M^{\mu\nu},R_{FK}\right],R_{FK}\right]\right\}\left|p_1,\ldots,p_f\right\rangle=-\Phi^{\mu\nu}_{LG}\left|p_1,\ldots,p_f\right\rangle\,.
\end{eqnarray}
Gathering this contribution and the one from $\Delta\Phi^{\mu\nu}_{FK}$, we finally arrive at \eqref{eq:toshow3} as required. This completes our proof that the dressed states of QEMD transform with exactly the same pairwise little group phase as the pairwise states defined in Section~\ref{sec:pairwisereview}.

\section{The Geometric Phase of Dressed-Pairwise States}

In the previous sections we proved that the transformation law for pairwise multiparticle states \eqref{eq:2particle} coincides with the one for the dressed multiparticle states of QEMD, \eqref{eq:dressedEM}. In this section we elaborate more on the geometric aspects of this correspondence. A key element in our derivation of both the pairwise states and the dressed states was the emergence of a \textit{geometric phase}, or Berry phase. This shouldn't come as a surprise, since after all the Aharonov-Bohm phase \cite{Aharonov1959} for a charge encircling a magnetic flux is a quintessential example of a geometric phase.

To see the geometric phase for the pairwise/dressed states, consider a rotation of the Dirac string,
\begin{eqnarray}\label{eq:Berry1}
n^\mu(\tau)=\exp\left[\tau\omega\right]^\mu_\nu\,n^\nu_0\,,
\end{eqnarray}
where $\tau$ parametrizes the amount of rotation.
As the Dirac string rotates, we have
 \begin{eqnarray}\label{eq:Berry2}
\left|p_1,\ldots,p_f\right\rrangle_{n(\tau+\delta \tau)}=e^{-\frac{i\delta\tau}{2}\omega_{\mu\nu} \Phi^{\mu\nu}_{LG}}\left|p_1,\ldots,p_f\right\rrangle_{n(\tau)}\,,
\end{eqnarray}
where $\Phi^{\mu\nu}_{LG}$ is given in \eqref{eq:PhimunuLG}.
Consequently
 \begin{eqnarray}\label{eq:Berry3}
\frac{d}{d\tau}\left|p_1,\ldots,p_f\right\rrangle=-\frac{i}{2}\omega_{\mu\nu} \Phi^{\mu\nu}_{LG}\left|p_1,\ldots,p_f\right\rrangle\,.
\end{eqnarray}
The geometric phase of the system is then given by \cite{Berry1984}
 \begin{eqnarray}\label{eq:Berry4}
\gamma_{Berry}&=&i\int_{0}^{2\pi}\,d\tau\,\left\llangle p_1,\ldots,p_f\right|\frac{d}{d\tau}\left|p_1,\ldots,p_f\right\rrangle=\frac{\omega_{\mu\nu}}{2} \int_{0}^{2\pi}\,d\tau\,\Phi^{\mu\nu}_{LG}\nonumber\\[5pt]
&=&\sum_{l<m}\,q_{lm}\,\int_{0}^{2\pi}\,d\tau\,\frac{\tau_{lm}\,n^{\mu}(\tau)\omega_{\mu\nu}\epsilon^{\nu}\left[p_l,p_m,n(\tau)\right]}{\epsilon^2\left[p_l,p_m,n(\tau)\right]}\nonumber\\[5pt]
&=&\sum_{l<m}\,q_{lm}\,\int_{0}^{2\pi}\,d\tau\,\frac{\tau_{lm}\,n^{\mu}_0\omega_{\mu\nu}\epsilon^{\nu}\left[p_l(\tau),p_m(\tau),n_0\right]}{\epsilon^2\left[p_l(\tau),p_m(\tau),n_0\right]}\nonumber\\
\end{eqnarray}
where $p_i(\tau)=\exp\left[-\tau\omega\right]^\mu_\nu\,p^\nu_i$. Straightforward integration gives
 \begin{eqnarray}\label{eq:Berry5}
\gamma_{Berry}&=&\pm 2\pi \sum_{l<m}\,q_{lm}\,.
\end{eqnarray}
We see that the system indeed has a geometric phase related to a rotation of the Dirac string, or conversely an inverse rotation of the momenta. To reproduce Dirac quantization, note that a geometric phase of $2n\pi$ means that our multiparticle state is bosonic, while a phase of $(2n+1)\pi$ means that our state is fermionic. Demanding that the overall multiparticle state is either a boson or a fermion, we get Dirac quantization, $q_{lm}=n/2$.  Interestingly, the geometric phase is independent of the direction of the string even if Dirac quantization does not hold; instead, Dirac quantization emerges from our rejection of fractional statistics.

Lastly, we comment on the difference between the usual Aharonov-Bohm argument for Dirac quantization and our geometric phase argument. In the standard argument, the charge is taken in a closed orbit around the string and picks up an Aharonov-Bohm phase proportional to the string's magnetic flux, $\gamma_{AB}=4\pi q$. The demand that the string is not observable leads to half-integer Dirac quantization, which guarantees that $\gamma_{AB}\sim0$ (mod $2 \pi$). In contrast, our geometric phase is string independent, and so it is allowed to be nontrivial, i.e. $\gamma_{Berry}\sim\pi$, if $\sum q$ is a half-integer. In this case the overall spin-statistics of our dressed/pairwise state is flipped. In \cite{Schwinger1975,Hasenfratz1976,Goldhaber1976,Brandt1978,Wilczek1982,Wilczek1982a}.
it was shown that this is completely consistent with the spin-statistics theorem. Here we provided a complete quantum field theoretic derivation of this fact, and also explained its origin in the soft photons exchanged between the charge and the monopole.

\section{Conclusions}\label{sec:conc}

In this paper we have unified many disparate concepts in the definition of quantum multiparticle states of charges and monopoles.
In particular, we proved that the pairwise multiparticle states, previously defined using group theory alone, coincide with the soft-photon dressed states of QEMD. To show this, we explicitly evaluated the action of the Noether generator $M^{\mu\nu}$ for Lorentz transformations on the dressed states of QEMD, and showed that they transform with exactly the same phase as predicted by the pairwise little group for pairwise states. 

As a byproduct of our work, we've shown that the $\mathcal{O}(eg)$ contribution to the soft-photon phase $\Phi_{FK}$ is \textit{finite} and has a geometric interpretation as a dihedral angle between two 3-planes in 4D. Moreover, $\Phi_{FK}$ plays a key role in setting the Lorentz transformation properties of QEMD dressed states. This is in contrast with $\Phi_{FK}$ in QED which is log divergent but is usually ignored, and which has no associated little group phase.

In the last part of the paper, we showed how the pairwise little group phase of the pairwise/dressed states leads to a \textit{geometric phase} when the Dirac string undergoes a full $2\pi$ rotation. This phase is independent of the string direction and equal to half of the familiar Aharonov-Bohm phase associated with encircling the Dirac string. Requiring the geometric phase to be a multiple of $\pi$ leads to half-integer quantization of $q_{ij}=(e_ig_j-e_jg_i)/4\pi$. Remarkably, for $\sum q_{ij}$ half-integer, the geometric phase is $\pi$, and so the overall pairwise / dressed state acquires opposite spin-statistics. A dressed state of one scalar monopole and one scalar charge with half integer $\sum q_{ij}$, for example, transforms as a fermion due to the photon coherent state sourced by the two mutually non-local charges. This effect, which has been previously discussed in the background-monopole limit, is now demonstrated for the first time in a complete QFT setting.

Finally, we wish to comment on two interesting future directions.   The first is a generalization to mutually non-local charged objects in different dimensions---for example $p$-branes and $d-p-4$ branes in $d$ dimensions. In particular, in 3D the monopole becomes an instanton connected to a string. The world line of a charge has a linking number with the string from the monopole/instanton, and we can get Dirac quantization. Repeating the calculation of the pairwise and geometric phases in this paper in a 3D setup, we expect to reproduce fractional statistics for anyons in 3D. Another future direction in 4D is to study the interplay of our QEMD dressing with the subleading soft photon theorem and with the subleading asymptotic Ward identity in QEMD. Our conjecture is that these would directly generalize to QEMD by the replacement $M^{\mu\nu}_{QED}\rightarrow M^{\mu\nu}_{QEMD}$.

\section*{Acknowledgements}
YS thanks Amit Sever for helpful discussions. CC and OT thank the hospitality of the Aspen Center for Physics, which is supported by NSF grant PHY-1066293. CC and ZD are supported in part by the NSF grant PHY-2014071. CC is also supported in part by the BSF grant 2020220. The work of SY was supported by a center of excellence of the Israel Science Foundation (grant number 2289/18). J.T. is supported by the DOE under grant  DE-SC-0009999. OT is supported in part by the DOE under grant DE-AC02-05CH11231.

\appendix

\section{Cancellation of Redundant Integrals}\label{app:red}
By \eqref{eq:DLI2def} we have
\begin{eqnarray}\label{eq:I11apB}
\Delta\mathcal{I}_{11}&=&p^\mu_1\mathcal{J}_\mu\nonumber\\[5pt]
\mathcal{J}^\mu&\equiv&\int\frac{d^4k}{(2\pi)^4}\frac{k^\mu}{(k^2+i\epsilon)(p_2\cdot k+i\epsilon)(p_3\cdot k+i\epsilon)(p_4\cdot k+i\epsilon)}\,.
\end{eqnarray}
By Lorentz invariance, we can cast $\mathcal{J}^\mu$ in the form 
\begin{eqnarray}\label{eq:projapB}
\mathcal{J}^\mu&=&C_2\,p^\mu_2\,+\,C_3\,p^\mu_3\,+\,C_4\,p^\mu_4\,,
\end{eqnarray}
where
\begin{eqnarray}
\colvec{I_{34}\\I_{24}\\I_{23}}=\colvec{p_2\cdot\mathcal{J}\\p_3\cdot\mathcal{J}\\p_4\cdot\mathcal{J}}=\colmatth{m^2_2& p_{23}&p_{24}\\p_{23}& m^2_{3}&p_{34}\\p_{24}& p_{34}&m^2_4}\,\colvec{C_2\\C_3\\C_4}\,,
\end{eqnarray}
and so
\begin{eqnarray}
\colvec{C_2\\C_3\\C_4}=\colmatth{m^2_2& p_{23}&p_{24}\\p_{23}& m^2_{3}&p_{34}\\p_{24}& p_{34}&m^2_4}^{-1}\,\colvec{I_{34}\\I_{24}\\I_{23}}\,.
\end{eqnarray}
Substituting \eqref{eq:projapB} in \eqref{eq:I11apB}, we have
\begin{eqnarray}\label{eq:projapBnew}
\Delta\mathcal{I}_{11}&=&C_2\,p_{12}\,+\,C_3\,p_{31}\,+\,C_4\,p_{14}\nonumber\\[5pt]
&=&\left(p_{12},\,p_{31},\,p_{14}\right)\colmatth{m^2_2& p_{23}&p_{24}\\p_{23}& m^2_{3}&p_{34}\\p_{24}& p_{34}&m^2_4}^{-1}\,\colvec{I_{34}\\I_{24}\\I_{23}}\nonumber\\[5pt]
&=&-\frac{f_{12}I_{34}+f_{31}I_{24}+f_{24}I_{31}}{f_{11}}\,.
\end{eqnarray}
A similar expression holds for $\Delta \mathcal{I}_{22}$.

\section{Miscellaneous Integrals}\label{app:misc}
\addtocontents{toc}{\protect\setcounter{tocdepth}{1}}
\subsection{$\Delta\mathcal{I}_{**}$}\label{app:miscIstst}
The integral $\Delta\mathcal{I}_{**}$ is given by
\begin{eqnarray}\label{eq:DLI4app}
\Delta\mathcal{I}_{**}&=&-i\bar{\epsilon}\int\frac{d^4k}{(2\pi)^2}\,\frac{\delta\left(p_1\cdot k\right)\,\delta\left(p_2\cdot k\right)}{(p_3\cdot k+i\epsilon)(p_4\cdot k+i\epsilon)}\,.
\end{eqnarray}
We can calculate this integral more easily by choosing a particular reference frame and then uplifting the result to a fully covariant expression. For our purposes, we can choose to work in the $(1,2)$ COM frame so that $p_1=(E_1,p\hat{z})$ and $p_2=(E_2,-p\hat{z})$. In this frame we can easily fix $k^t$ and $k^z$ by integrating over the delta functions. We have
\begin{eqnarray}\label{eq:DLI4app2}
\Delta\mathcal{I}_{**}&=&i\frac{\bar{\epsilon}}{p\,(E_1+E_2)}\int\frac{d^2k}{(2\pi)^2}\,\frac{1}{(\vec{p}_3\cdot \vec{k}+i\epsilon)(\vec{p}_4\cdot \vec{k}+i\epsilon)}\nonumber\\[5pt]
&=&i(p^x_3p^y_4-p^y_3p^x_4)\int\frac{d^2k}{(2\pi)^2}\,\frac{1}{(\vec{p}_3\cdot \vec{k}+i\epsilon)(\vec{p}_4\cdot \vec{k}+i\epsilon)}\,,
\end{eqnarray}
where $\vec{p}_{3,4}=\left(p^x_{3,4},p^y_{3,4}\right)$. The second equality here stems from $\bar{\epsilon}=p\,(E_1+E_2)(p^x_3p^y_4-p^y_3p^x_4)$. Using Schwinger parameters, we have
\begin{eqnarray}\label{eq:DLI4app3}
\Delta\mathcal{I}_{**}&=&i(p^x_3p^y_4-p^y_3p^x_4)\,\int_0^\infty\,d\alpha_3\,\int_0^\infty\,d\alpha_4\,\int\frac{d^2k}{(2\pi)^2}\,e^{i\vec{k}\cdot(\alpha_3 \vec{p}_3+\alpha_4\vec{p}_4)}\nonumber\\[5pt]
&=&i(p^x_3p^y_4-p^y_3p^x_4)\,\int_0^\infty\,d\alpha_3\,\int_0^\infty\,d\alpha_4\,\delta^{(2)}\left[\alpha_3 \vec{p}_3+\alpha_4 \vec{p}_4\right]\,.
\end{eqnarray}
Changing variables as $\vec{z}=\alpha_3 \vec{p}_3+\alpha_4 \vec{p}_4$, we have
\begin{eqnarray}\label{eq:DLI4app4}
\Delta\mathcal{I}_{**}&=&i\iint_{A}\,\delta^{(2)}\left(\vec{z}\right)\,,
\end{eqnarray}
Where $A=\left\{\vec{z}\,|\,\vec{z}=\alpha_3 \vec{p}_3+\alpha_4 \vec{p}_4,0\leq\alpha_3,0\leq\alpha_4\right\}$. Clearly, the integral over the delta function picks us the part of the 2D plane spanned by linear combinations of $\vec{p}_3$ and $\vec{p}_4$ with positive coefficients, and so 
\begin{eqnarray}\label{eq:DLI4app5}
\Delta\mathcal{I}_{**}&=&\frac{i}{2\pi}\,\arccos\left(\frac{\vec{p}_3\cdot\vec{p}_4}{|\vec{p}_3||\vec{p}_4|}\right)\,.
\end{eqnarray}
Uplifting this to a fully covariant expression, we have
\begin{eqnarray}\label{eq:DLI4app6}
\Delta\mathcal{I}_{**}&=&\frac{i}{2\pi}\,\arccos\left[\hat{\epsilon}(p_1,p_2,p_3)\cdot\hat{\epsilon}(p_1,p_2,p_4)\right]\,.
\end{eqnarray}
\subsection{$G(z)$}\label{app:miscGz}
For future reference we define the 4D Green's function
\begin{eqnarray}
G(z)=\int \frac{d^{3} k}{(2 \pi)^{3}} \frac{1}{2 \omega_k}\,e^{-ik\cdot z}\,.
\end{eqnarray}
By the residue theorem, we have
\begin{eqnarray}
G(z)=\lim_{\mu\rightarrow 0}\int \frac{d^{4} k}{(2 \pi)^{4}} \frac{i}{k^2-\mu^2+i\epsilon}\,e^{-ik\cdot z}\,,
\end{eqnarray}
where we put a regulator mass that we later take to $0$. Using a Schwinger parameter, we have
\begin{eqnarray}\label{eq:G}
G(z)&=&\lim_{\mu\rightarrow 0}\int_0^\infty d\alpha\,\int \frac{d^{4} k}{(2 \pi)^{4}}e^{i(\alpha k^2-\alpha \mu^2-k\cdot z)}\nonumber\\[5pt]
&=&-\frac{i}{16\pi^2}\lim_{\mu\rightarrow 0}\int_0^\infty d\alpha\,\frac{1}{\alpha^2}e^{i(\frac{z^2}{4\alpha}+\alpha \mu^2)}\nonumber\\[5pt]
&=&\frac{1}{4\pi^2}\lim_{\mu\rightarrow 0}\frac{-i\mu|z|K_1(i\mu|z|)}{z^2+i\epsilon}=\frac{1}{4\pi^2}\frac{1}{z^2+i\epsilon}\,.
\end{eqnarray}
\subsection{$\mathcal{I}_{A}$}\label{app:miscIA}
The $\mathcal{I}_{A}$ integral of section~\ref{sec:IA} is given by:
\begin{eqnarray}
\mathcal{I}_{A}(p_1,p_2,p_3)=-\frac{8\pi t}{\omega_1}\,\int_{-\infty}^\infty\,\frac{dt'}{\omega_2}\,\,\text{Im}\left\{\int \frac{d^{3} k}{(2 \pi)^{3}} \frac{1}{2 \omega_k}\,\frac{p^{[\mu}_1\epsilon^{\nu]}(p_2,p_3,k)}{p_3\cdot k +i\epsilon}e^{-ik\cdot\left(\frac{p^\mu_1t}{\omega_1}-\frac{p^\mu_2t'}{\omega_2}\right)}\right\}\,,\nonumber\\
\end{eqnarray}
where we again defined $p_3\equiv n$. Evaluating it using a Schwinger parametrization, we get
\begin{eqnarray}
\mathcal{I}_{A}(p_1,p_2,p_3)&=&-\frac{8\pi t}{\omega_1}\,\text{Im}\left\{\int_0^\infty d\alpha\int_{-\infty}^\infty\,\frac{dt'}{\omega_2}~p^{[\mu}_1\epsilon^{\nu]}(p_2,p_3,\partial_{z})\,G(z)|_{z=z_*}\right\}\,,
\end{eqnarray}
where $z=\frac{p^\mu_1t}{\omega_1}-\frac{p^\mu_2t'}{\omega_2}-\alpha p_3$.
Substituting $G(z)$ from \eqref{eq:G}, we have
\begin{eqnarray}
\mathcal{I}_{A}(p_1,p_2,p_3)&=&-\frac{t^2}{\omega^2_1}\,\frac{2\,p^{[\mu}_1\epsilon^{\nu]}\left(p_1,p_2,p_3\right)}{\pi}\,\text{Im}\left\{\int_0^\infty d\alpha\,\int_{-\infty}^\infty\,\frac{dt'}{\omega_2}\,\frac{1}{(p_1t/\omega_1-p_2t'/\omega_2-\alpha p_3)^4}\right\}\nonumber\\[5pt]
&=&-\frac{t^2}{\omega^2_1}\,p^{[\mu}_1\epsilon^{\nu]}\left(p_1,p_2,p_3\right)\,\text{Re}\left\{\int_0^\infty d\alpha\,\frac{m^2_2}{{\left[((p_1t/\omega_1-\alpha p_3)\cdot p_2)^2-m^2_2(p_1t/\omega_1-\alpha p_3)^2\right]}^{\frac{3}{2}}}\right\}\nonumber\\[5pt]
&=&\frac{p^{[\mu}_1\epsilon^{\nu]}\left(p_1,p_2,p_3\right)}{\tau_{12}}\,\frac{m^2_2\,p_{31}-p_{12}p_{23}}{\epsilon^2(p_1,p_2,p_3)}\,.
\end{eqnarray}
\subsection{$\mathcal{I}_{F,l}$}\label{app:miscIFl}
The $\mathcal{I}_{F,l}$ integral of section~\ref{sec:IF} is given by
\begin{eqnarray}\label{eq:EMangcom1app}
\mathcal{I}_{F,l}(p)&=&-2i\int_{-\infty}^\infty dt'~\text{Re}\left\{\int \frac{d^{3} k}{(2 \pi)^{3}} \frac{1}{2 \omega_k}~\left[e_lk^{[\mu}p^{\nu]}-g_l\epsilon^{\mu\nu}\left(k,p\right) \right]e^{-ik\cdot(x-\frac{p}{\omega_p} t')}\right\}\,.
\end{eqnarray}
We can present $\mathcal{I}_{F,l}(p)$ in the form
\begin{eqnarray}\label{eq:EMangcom1app2}
\mathcal{I}_{F,l}(p)&=&-2i\int_{-\infty}^\infty dt'~\text{Re}\left\{e_l\,I^{[\mu}_F(p)p^{\nu]}-g_l\,\epsilon^{\mu\nu}\left(I_F(p),p\right) \right\}\,\nonumber\\[5pt]
I^\mu_{F}(p)&\equiv& \int \frac{d^{3} k}{(2 \pi)^{3}} \frac{k^{\mu}}{2 \omega_k}\,e^{-ik\cdot(x-\frac{p}{\omega_p} t')}=i\frac{\partial}{\partial x_\mu}\,G\left(x-\frac{p}{\omega_p} t'\right)\,.
\end{eqnarray}
Substituting $G(z)$ from \eqref{eq:G}, we have
\begin{eqnarray}\label{eq:EMangcom15app}
\mathcal{I}_{F,l}(p)&=&-\frac{i}{2\pi^2}\,\int_{-\infty}^\infty \frac{dt'}{\omega_p}~\,\frac{e_l\,x^{[\mu}p^{\nu]}-g_l\,\epsilon^{\mu\nu}\left(x,p\right)}{\left(x-\frac{p}{\omega_p} t'\right)^4}=-\frac{m^2}{4\pi}\,\frac{e_l\,x^{[\mu}p^{\nu]}-g_l\,\epsilon^{\mu\nu}\left(x,p\right)}{{\left[(p\cdot x)^2-m^2x^2\right]}^{\frac{3}{2}}}\,.\nonumber\\
\end{eqnarray}
\subsection{$\mathcal{I}^{\mu\nu}_{EM}$}\label{app:miscIEM}
The $\mathcal{I}^{\mu\nu}_{EM}$ integral of section~\ref{sec:IF} is given by
\begin{eqnarray}
\mathcal{I}^{\mu\nu}_{EM}(p_a,p_b)&=&\frac{m^2_am^2_bt}{4\pi}\int d^3x\,\frac{x^{[\mu}\epsilon^{\nu]}(x,p_a,p_b)}{{\left[(p_a\cdot x)^2-m^2_ax^2\right]}^{\frac{3}{2}}{\left[(p_b\cdot x)^2-m^2_bx^2\right]}^{\frac{3}{2}}}\,.
\end{eqnarray}
To calculate it, we note that its antisymmetry means that it is spanned by $p^{[\mu}_1p^{\nu]}_2$ and $\epsilon^{\mu\nu}(p_1,p_2)$. We can check explicitly that its contraction with the former vanishes, and so it is only proportional to the latter. Projecting along $\epsilon^{\mu\nu}(p_1,p_2)$, we can relate $\mathcal{I}^{\mu\nu}_{EM}$ to a scalar integral as
\begin{eqnarray}\label{eq:EMangs2}
\mathcal{I}^{\mu\nu}_{EM}&=&-\frac{\mathcal{I}_{EM}}{\epsilon_{\alpha\beta}(p_a,p_b)\,\epsilon^{\alpha\beta}(p_a,p_b)}\epsilon^{\mu\nu}(p_a,p_b)=\frac{\mathcal{I}_{EM}}{2\tau^2_{ab}}\epsilon^{\mu\nu}(p_a,p_b)\nonumber\\[5pt]
\mathcal{I}_{EM}&\equiv&-\epsilon_{\alpha\beta}(p_a,p_b)\,\mathcal{I}^{\alpha\beta}_{EM}=\frac{2m^2_am^2_bt}{4\pi}\int d^3x\,\frac{-x^2\tau^2_{ab}+2(x\cdot p_a)(x\cdot p_b)p_{ab}-m^2_a(x\cdot p_b)^2-m^2_b(x\cdot p_a)^2}{{\left[(p_a\cdot x)^2-m^2_ax^2\right]}^{\frac{3}{2}}{\left[(p_b\cdot x)^2-m^2_bx^2\right]}^{\frac{3}{2}}}\,.\nonumber\\
\end{eqnarray}
To calculate $\mathcal{I}^{\mu\nu}_{EM}$, we first change variables as $x=(t,t\vec{y})$ and arrive at
\begin{eqnarray}\label{eq:EMangs3}
&&\mathcal{I}_{EM}=\frac{2m^2_am^2_b}{4\pi}\times\nonumber\\[5pt]
&&\int d^3y\,\frac{(|\vec{y}|^2-1)\tau^2_{ab}+2(\omega_a-\vec{y}\cdot\vec{p}_a)(\omega_b-\vec{y}\cdot\vec{p}_b)p_{ab}-m^2_a(\omega_b-\vec{y}\cdot\vec{p}_b)^2-m^2_b(\omega_a-\vec{y}\cdot\vec{p}_b)^2}{{\left[(\omega_a-\vec{y}\cdot\vec{p}_a)^2-m^2_a(1-|\vec{y}|^2)\right]}^{\frac{3}{2}}{\left[(\omega_b-\vec{y}\cdot\vec{p}_b)^2-m^2_b(1-|\vec{y}|^2)\right]}^{\frac{3}{2}}}\,.\nonumber\\
\end{eqnarray}
Since this is a scalar integral, we calculate it in a particular reference frame, and then ``uplift" the result to its unique covariant form. This frame is the COM frame with $-\vec{p_b}=\vec{p}_a=p\hat{z}$, $\omega_i=\sqrt{p^2+m^2_i}$. Using cylindrical coordinates for the integral, we have
\begin{eqnarray}\label{eq:EMangs4}
\mathcal{I}_{EM}&=&m^2_am^2_b\tau^2_{ab} \int_{-\infty}^{\infty} dz\,\int_0^\infty dr\,\frac{r^3}{{\left[m^2_ar^2+(\omega_az-p)^2\right]}^{\frac{3}{2}}{\left[m^2_br^2+(\omega_bz+p)^2\right]}^{\frac{3}{2}}}=2\tau_{ab} \,,\nonumber\\
\end{eqnarray}
and so
\begin{eqnarray}\label{eq:EMangsf}
\mathcal{I}^{\mu\nu}_{EM}&=&\frac{\epsilon^{\mu\nu}(p_a,p_b)}{\tau_{ab}}\,.
\end{eqnarray}

\bibliographystyle{JHEP}
\bibliography{AMP}{}
\end{document}